\documentclass[12pt]{iopart}
\usepackage{color}

\usepackage{iopams}  
\usepackage{graphicx}
\usepackage{wasysym}
\usepackage{cite}

\begin{document}

\title{Searching for new physics using optically levitated sensors}

\author{David C. Moore}
\address{Wright Laboratory, Department of Physics, Yale University, New Haven, CT, USA}
\ead{david.c.moore@yale.edu}

\author{Andrew A. Geraci}
\address{Center for Fundamental Physics, Department of Physics and Astronomy,
Northwestern University, Evanston, IL, USA}
\ead{andrew.geraci@northwestern.edu}

\vspace{10pt}

\begin{abstract}
We describe a variety of searches for new physics beyond the Standard Model of particle physics which may be enabled in the coming years by the use of optically levitated masses in high vacuum.  Such systems are expected to reach force and acceleration sensitivities approaching (and possibly eventually exceeding) the standard quantum limit over the next decade.  For new forces or phenomena that couple to mass, high precision sensing using objects with masses in the fg--ng range have significant discovery potential for new physics. Such applications include tests of fundamental force laws, searches for non-neutrality of matter, high-frequency gravitational wave detectors, dark matter searches, and tests of quantum foundations using massive objects.
\end{abstract}

\section{Introduction}

The Standard Model (SM) of particle physics is one of the most precisely tested theories developed to date~\cite{PDG2018}. For example, the predictions of quantum electrodynamics for the electron magnetic moment agree with experimental data at a precision exceeding 1 part per billion~\cite{Odom_g-2,PDG2018}. Despite this astounding success, there is a preponderance of evidence that the SM is incomplete. It fails to account for several basic phenomena of the universe: Why is there more matter than anti-matter in the universe? What is Dark Matter, which is five times more abundant than the ordinary matter described by the SM? What is Dark Energy? How can gravity be incorporated into our understanding of the other fundamental forces in the SM?  The traditional approach towards answering these questions has been to increase the collision energy achievable at particle colliders, allowing new particles and phenomena to be produced and studied. While this approach has been enormously successful during the development and confirmation of the SM, colliders have thus far failed to find any of the new physics Beyond the Standard Model (BSM) described above. As the cost of continuing to increase the energy frontier at colliders grows, searches at the precision frontier of particle physics provide an alternative approach~\cite{Safronova:2017xyt,Erler:2019_EW_precision,Jaeckel:2010ni}. At the precision frontier, sensitive laboratory-scale experiments may be able to provide insight into these questions by detecting tiny deviations arising from higher energy scales (or weaker couplings) than can currently be reached by collider experiments. 

Mechanical and opto-mechanical sensors such as torsion balances \cite{Kapner:2007, Yang2012, 2011PhRvL.106d1801H}, optical interferometers \cite{TheLIGOScientific:2014jea}, and microfabricated oscillators \cite{Deca:2016,Geraci:2008} have been a powerful tool in a variety of precision experiments, including searches for fifth-forces \cite{GiudiceDimopoulos,add,Deca:2016,Geraci:2008,adelberger2020,HUST:2020}, tests of the equivalence principle \cite{wagnerEPV}, and searches for gravitational waves \cite{LIGOfirst}. Gravitational wave observatories have already achieved quantum limited sensitivity over a part of their operating range, and  developing novel and improved quantum-limited and quantum-based sensors is a promising path towards extending the sensitivity frontier beyond the state of the art \cite{TheLIGOScientific:2014jea}.

\section{Overview of optically levitated systems}
Following pioneering work by Ashkin and Dziedzic~\cite{Ashkin:1971,Ashkin:1976,Ashkin:1977}, levitated optomechanics has undergone substantial development in the past decade~\cite{Millen:2020review}, with techniques now demonstrated to trap objects with diameters between $\sim$50~nm--10~$\mu$m using optical~\cite{Li:2011,Gieseler:2012,Moore:2014,Ranjit:2015,Pettit:2019,Huizhu:18,Vovrosh:2017,Aspelmeyer:2019}, magnetic~\cite{DUrso:2018,Ulbricht:2019,BrianD'Urso_2020,Gieseler:2020}, or RF~\cite{Dania:2020kzl,Bullier2020,Millen:2018} trapping fields.  Due to the high isolation from thermal and environmental sources of noise possible in a high-vacuum environment, such objects have found applications to precise force sensing and accelerometry~\cite{Ranjit:2015,Ranjit:2016,Acceleration_2017,acceleration2020,hempston2017force,Novotny_static:2018}, torque sensing~\cite{Hoang:2016,Tongcang_rotation:2018,Millen2017_2,vanderLaan2020}, electric field sensing~\cite{Blakemore_3D_microscope:2019,Hempston:2017}, and pressure sensing~\cite{blakemore_gauge:2019}.  In comparison to other high-sensitivity systems (e.g., torsion balances~\cite{EotWash2020}, atom interferometers~\cite{Asenbaum:2020era}, or ions~\cite{Gilmore:2017cbw,Biercuk:2010}), levitated nano- or micro-scale objects can have advantages when the forces or accelerations of interest arise at the $\mu$m (or sub-$\mu$m scale) and couple to mass, neutron number, or other quantities for which use of a macroscopic mass rather than single atom or ion is beneficial. These features make levitated optomechanical systems useful for probing a variety of BSM models in which weakly coupled, $\mu$m-scale interactions arise.

Here we focus on a specific subset of levitated optomechanical systems, i.e. optically levitated systems, and their possible sensitivity to BSM physics in the coming years.  Among the techniques employed in levitated optomechanics, optical levitation presents advantages for certain applications.  First, no net electric charge or magnetization is required to stably trap the object, minimizing coupling to backgrounds arising from stray electric or magnetic fields.  In addition, a variety of tools developed for manipulating optical potentials can be directly applied. These tools include the high-bandwidth control of the trapping lasers; optics that permit long working distances ($\gtrsim$ several cm) between the optical elements and the trapped objects to minimize coupling to nearby surfaces and permit electrodes and attractors to be positioned around the trap; and the straightforward ability to form multiple traps and eventually large arrays of levitated sensors~\cite{Carney_white_paper,carney:2019_1}.

\begin{figure}[t]
    \centering
    \includegraphics[width=\columnwidth]{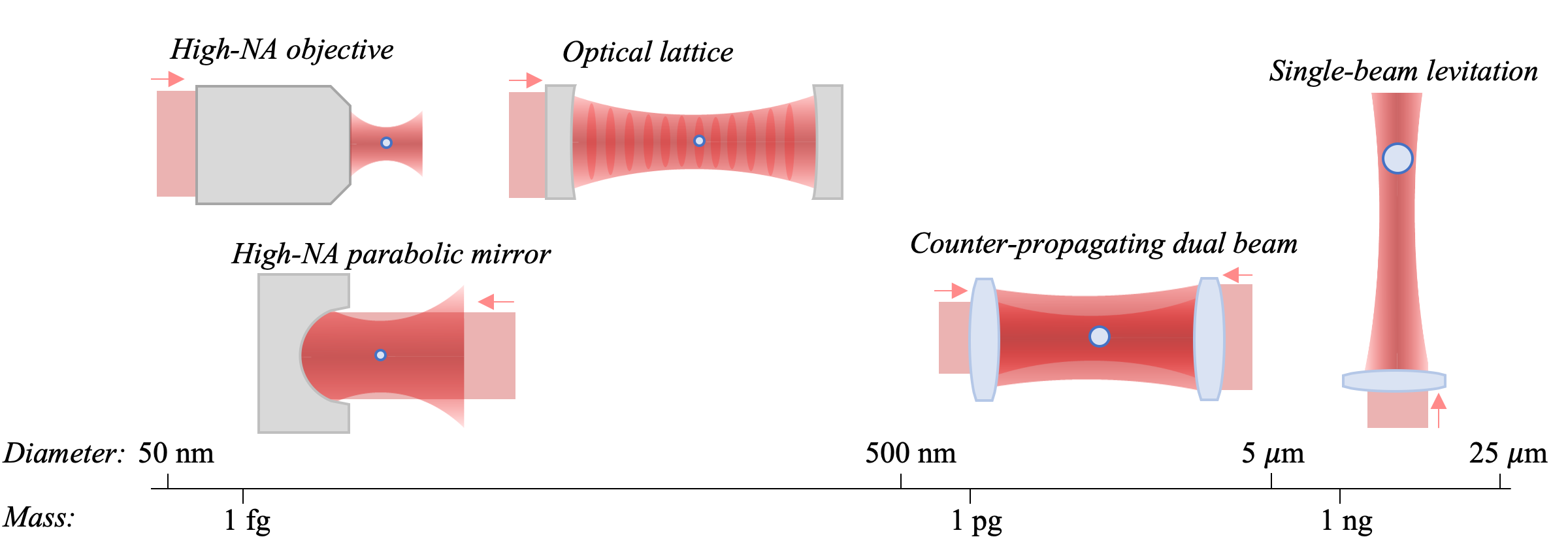}
    \caption{ Schematics of the various optical trapping geometries demonstrated in vacuum, for objects spanning more than 6 orders of magnitude in mass.  Subwavelength objects with masses $\sim 1-10$~fg have been trapped in high-NA single beams~\cite{Gieseler:2012,Vovrosh:2017} or optical lattices~\cite{Ranjit:2016}.  Objects with masses $\sim 1-10$~pg have been trapped in counter propagating dual-beam traps~\cite{Li:2011,Ranjit:2015}.  Objects with masses $\sim 1-10$~ng are typically trapped in single-beam levitation traps~\cite{Moore:2014,Acceleration_2017}. }
    \label{fig:trapping_setups}
\end{figure}

A variety of techniques for optical trapping in high vacuum have recently been demonstrated.  For objects with diameters, $d_{sph} \approx 50-300$~nm, traps employing tightly focused beams (NA $\gtrsim 0.9$) are typically required to provide sufficiently deep trapping potentials, and have been implemented either using microscope objectives~\cite{Gieseler:2012} or parabolic mirrors~\cite{Vovrosh:2017}.  Such sub-wavelength objects can also be trapped using standing wave potentials~\cite{Ranjit:2016}.  For objects with diameters between $1 \lesssim d_{sph} \lesssim 5\ \mu$m, both counter-propagating dual-beam traps~\cite{Li:2011,Ranjit:2015} as well as single-beam levitation traps~\cite{Moore:2014} have been demonstrated. At larger diameters, optical absorption and heating limit the maximum particle size for typical materials, and only single-beam levitation traps have been demonstrated in vacuum~\cite{Ashkin:1971,Acceleration_2017}.  Lower absorption materials may allow larger particles to be optically levitated~\cite{Ashkin:1976,acceleration2020}.  For example, assuming the typical absorption for optical quality fused silica, it should be possible to optically levitate particles as large as several hundred $\mu$m in diameter (using $\sim$1~W of laser power)~\cite{Acceleration_2017}.  For smaller (sub-wavelength) particles, the trap depth decreases linearly with the sphere volume, leading to practical limitations to sphere sizes $\gtrsim 20$~nm when trapping nanospheres at room temperature in most applications to date~\cite{Millen:2020review}.  Smaller particles can be trapped if pre-cooled to sufficiently low temperature prior to loading into the trap (including, ultimately, single atoms at $\lesssim$mK temperatures)~\cite{2000AAMOP..42...95G}.

A key advantage of optically levitated sensors for the applications considered here is the ability to trap neutral objects (or neutralize any net charge present on the object).  Several techniques have been demonstrated for {\em in situ} control of the charge state of an optically trapped microsphere or nanosphere, either by removing or adding electrons in controlled steps of $\pm 1\ e$~\cite{Moore:2014,acceleration2020,Frimmer:2017,Conangla:2018nnn}.  While $\mu$m-sized objects are typically trapped with a substantial net charge (as large as $\sim 10^4\ e$) after loading~\cite{acceleration2020}, sub-$\mu$m objects have a significant probability to be trapped with a net charge of $0~e$~\cite{Frimmer:2017,Ranjit:2016}.  Once neutral, optically trapped objects have been observed to remain neutralized (without spontaneous charging) for time periods of a week or longer~\cite{acceleration2020}.

\section{Progress towards the quantum measurement regime}
A key feature of levitated optomechanical systems is their ability to evade thermal noise from direct contact with the environment---which is always present in clamped oscillators---simply by operating at ultra-high vacuum, but while the vacuum chamber walls remain at room temperature.  At pressures below $10^{-8} - 10^{-10}$~mbar, depending on mass, other sources of noise become dominant over thermal noise~\cite{Millen:2020review}.  In the absence of additional technical noise, the sensitivity for measuring the sphere displacement is limited by photon shot noise from the trapping light, with wavelength $\lambda$.  This shot noise arises both from the random arrival time of photons in the trapping beam, and includes impulses arising from the random scattering of the photons by the sphere (i.e., ``photon recoil heating'')~\cite{Chang:2010,Jain:2016}.  We note that the detailed rate of photon recoil heating is in general dependent on the geometry of the object:  for sub-wavelength objects that randomly scatter light as a point dipole (i.e., relatively isotropically), such heating due to the random direction of scatters is largely unavoidable.  In contrast, for larger objects that do not act as point dipoles, certain geometries can avoid the component of the photon recoil heating arising from the random direction of scattering~\cite{Aggarwal:2020umq,Chang_2012}, although more generally radiation pressure shot noise will still be present~\cite{2013Sci...339..801P,PhysRevD.23.1693}.  Finally, if only a fraction of the scattered photons are captured, there is additional shot noise in the detection of the sphere position.  Although in principle the detection process can be made highly efficient, maximizing photon detection efficiency is a key challenge for levitated optomechanical systems~\cite{Tebben2019}.

\begin{figure}[t]
    \centering
    \includegraphics[width=\columnwidth]{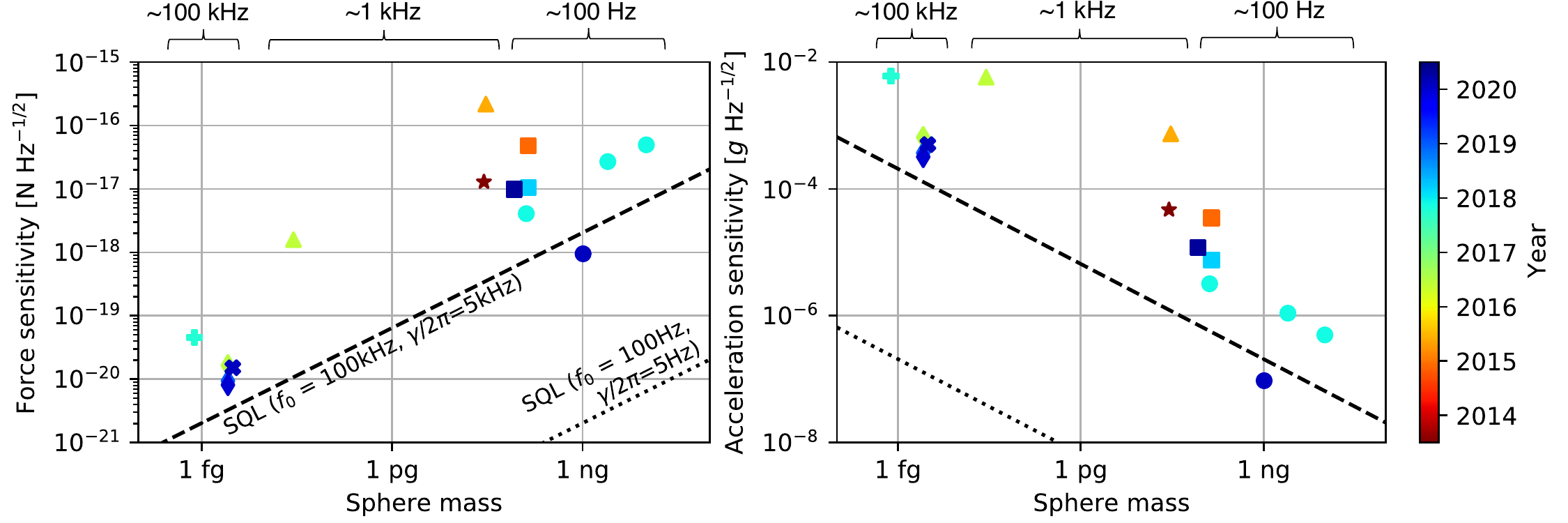}
    \caption{ Compilation of measured force and acceleration sensitivities for optically levitated masses range between $\sim 1$~fg to $\sim$10~ng from a variety of groups: ETH Zurich ($\blacklozenge$)~\cite{Novotny2019,Novotny:2020}, Southampton ($\boldsymbol{+}$)~\cite{Hempston:2017} Stanford ($\blacksquare$)~\cite{Moore:2014,Rider:2017,Kawasaki:2020oyl}, UN Reno/Northwestern ($\blacktriangle$)~\cite{Ranjit:2015,Ranjit:2016}, UT Austin ($\star$)~\cite{Li:2011}, Vienna ($\boldsymbol{\times}$)~\cite{Aspelmeyer:2019}, and Yale ($\CIRCLE$)~\cite{Acceleration_2017,acceleration2020}. All points directly correspond to reported force or acceleration sensitivities, with the exception of Refs~\cite{Novotny2019,Novotny:2020,Aspelmeyer:2018} where such sensitivities are not directly reported, and instead are estimated from the reported effective temperature and damping. The typical resonant frequencies of the traps varies significantly over the mass range considered and is denoted at the top of the plot.  Points are colored by time, showing continued improvement in sensitivities over time and progress towards reaching the relevant SQL in various mass ranges.  In addition to the data shown, substantial progress has also been made to develop ultra-sensitive torque sensors and free rotors~\cite{arita2016rotational,Millen2017_2,vanderLaan2020,Rotation_paper_2018,Rider:2019,Ahn:2020}, with sensitivities reaching $\sim10^{-27}$~N\,m~(Purdue~\cite{Ahn:2020}). }
    \label{fig:sens_vs_SQL}
\end{figure}

Assuming technical noise sources are mitigated and high collection efficiency of the scattered light, shot noise due to the random arrival times of photons can dominate the sensitivity at low laser powers, while measurement backaction processes (i.e., displacement of the sphere's COM due to shot noise in the rate of photon recoils) become dominant at high powers. At the ``standard quantum limit'' (SQL), the trapping power is set such that noise contributions from shot noise and backaction are equal, for a given measurement frequency. For force sensitivity, the on-resonant SQL can be estimated as $\sqrt{S_{FF}^{SQL}} = \sqrt{2\hbar m \omega_0 \gamma}$, where $m$ is the sphere mass, $\omega_0 = 2\pi f_0$ for sphere resonant frequency $f_0$, and $\gamma$ is the damping coefficient~\cite{Mason:2019piu,carney:2019_2}.  The resulting SQL for several benchmark parameters is shown in Fig.~\ref{fig:sens_vs_SQL}.  Note for systems approaching ground state cooling, where $k_b T \approx \hbar \omega_0$, this expression approximately reproduces the spectral density for a classical oscillator, $\sqrt{S_{FF}} = \sqrt{2 k_b T m \gamma}$~\cite{Kubo_1966}, where $k_b$ is Boltzmann's constant and $T$ is the temperature. While the SQL does not present a fundamental limit---since squeezing or backaction evasion can continue to be used to surpass the SQL~\cite{Mason:2019piu}---reaching such sensitivities is an important goal for a given class of optomechanical sensors.  Optically levitated $\sim$fg masses are already approaching the relevant SQL~\cite{Novotny:2020,Aspelmeyer:2019}, while technical noise in the displacement sensitivity for more massive objects is still $\gtrsim 10^2$ times above that needed to reach the SQL for the most sensitive systems demonstrated to date~\cite{acceleration2020}.  

In the coming years, reaching the SQL (and possibly surpassing it) for a range of masses of levitated objects will further enhance sensitivity to the searches for BSM physics described here, provided backgrounds can also be sufficiently mitigated.  For many of these searches, the acceleration or force sensitivity of a levitated sensor, rather than its displacement sensitivity, it typically of interest. A key advantage of levitated fg--ng masses is that they concentrate a substantial mass density within a sub-$\mu$m length scale, which can provide substantially better acceleration sensitivity than for single trapped ions or atoms, reaching $\sim$100~n$g$/$\mathrm{\sqrt{Hz}}$ for 10~$\mu$m diameter spheres~\cite{acceleration2020}.  In contrast, less massive objects reach better force sensitivities (e.g., $\sim$$10^{-18}\ \mathrm{N/\sqrt{Hz}}$ for microspheres, $\sim$$10^{-20}\ \mathrm{N/\sqrt{Hz}}$ for nanospheres, and $\sim$$10^{-24}\ \mathrm{N/\sqrt{Hz}}$ for ions), provided the sensor can be coupled to the force of interest.  For forces that couple to mass or number of nucleons, acceleration sensitivity is typically the appropriate figure of merit.  Also important is the length scale of interest for the accelerometer if looking for short range forces (see Sec.~\ref{sec:forces}) or the sensor mass if searching for impulses (see Sec.~\ref{sec:DM}).

Fig.~\ref{fig:sens_vs_SQL} shows a summary of the current force and acceleration sensitivity achieved with a variety of optically levitated systems.  Continued improvement over the last decade, since the first modern implementation of optical trapping and cooling in 2011~\cite{Li:2011} has led to 1--2 orders of magnitude improvement over masses ranging from fg--ng scales. 

While the above discussion has focused on displacement sensitivity (and, in particular, its translation to force/acceleration sensitivity), an additional benchmark for reaching the quantum measurement regime for optically levitated systems is to cool their center-of-mass motion to the ground state of the trapping potential.  Ground state cooling is typically a prerequisite for implementations of advanced measurement techniques requiring non-trivial state preparations of the massive oscillator~\cite{Oriol2011,Geraci:matter_wave,Ulbricht:2014,Marletto:2017,Bose:2017}. Recent work has demonstrated ground-state (or near ground-state) cooling of $\sim$fg scale optically levitated masses via either cavity cooling through coherent scattering~\cite{Aspelmeyer:2019} or feedback damping~\cite{Novotny2019,Novotny:2020}.  For more massive objects, technical noise still presents a substantial barrier to ground state cooling, although effective temperatures as low as $\sim$50~$\mu$K have now been reached for ng masses~\cite{acceleration2020}.  Many of the technical developments required to permit displacement sensitivity at the SQL would also translate into an improved ability to optically cool these more massive optically trapped objects. We note that in practice in the thermal regime, cavity-cooling~\cite{Chang:2010} and laser feedback cooling (e.g. cold damping)~\cite{Li:2011}, tend to reduce the center-of-mass motional temperature and increase the damping rate by a similar factor, thus leaving the thermally limited force sensitivity unaffected, while at the same time allowing a useful tuning of the effective sensor bandwidth.   While such cooling does not affect the sensitivity of the oscillator to continuous searches for forces or accelerations, it is a key figure-of-merit for their sensitivity to impulses such as those possibly resulting from dark matter described in Sec.~\ref{sec:DM}.  This is because impulse searches (unlike measurements of continuous forces) compare the instantaneous energy of the oscillator to its steady state temperature in the presence of cooling, requiring a sufficiently low initial temperature to observe a given impulse amplitude.  In general, the optimum cooling rate for feedback cooling in impulse sensing applications is the highest rate possible prior to the onset of noise squashing~\cite{Monteiro:2020wcb}.  

A final ambitious goal for the levitated optomechanics community would be to harness the effects of macroscopic quantum interference and superposition with optically levitated particles to perform matter wave interferometry. This has potential application for tests of force laws, tests of gravitational waves, as well as tests of quantum foundations \cite{Oriol2011}. Single-particle matter wave interference has been long established with electrons, neutrons, atoms, and molecules \cite{matterwavereview}. The largest object explicitly shown to obey wave-particle duality by diffracting it from a grating is a macromolecule with a mass of approximately $10^4$ amu \cite{arndt2019}. For a nanosphere with a sufficiently cooled center-of-mass, it should be possible to release the particle from the optical trap to obtain a point-like source for a matter wave interferometer \cite{Ulbricht:2014,Geraci:matter_wave} with a mass as high as $10^7$ amu. The wave packet of the nanoparticle will expand upon release from the trap, and can be diffracted from a material grating or a light-grating, in a similar fashion to what has been achieved for macromolecules \cite{arndt2019}.  After diffracting the particle the position can be recorded, and subsequent repeated experiments can be undertaken to construct an interference pattern, one particle at a time.  The location of the fringes of this interference pattern depends on the acceleration experienced by the particle along the direction of the grating, and the free-fall interferometer thus can be used as a sensitive accelerometer for precision fundamental physics tests \cite{Geraci:matter_wave}. As already demonstrated in free-fall atom interferometers, and acceleration sensitivity of $\sim \mathrm{n}g/\sqrt{{\rm{Hz}}}$ is also in principle possible in nanoparticle systems, but with a much-reduced wave-packet expansion when compared with an atomic interferometer due to the much larger mass. While the acceleration sensitivity may be comparable to what is achievable in atom interferometers, this reduced expansion could permit higher-spatial resolution and more localized sensing.

\section{Searches for new physics}
\subsection{Tests of force laws}
\label{sec:forces}
The acceleration and force sensitivity achievable with optically levitated objects can enable their use to test the fundamental forces laws for the known long-range interactions in the SM, i.e., Newton's and Coulomb's laws down to $\mu$m and sub-$\mu$m distances.  These tests may be sensitive either to modifications to the gravitational force law that appear only at short distances, or new ``fifth forces'' that couple to, e.g., mass or electric charge.  Such forces can appear in a variety of BSM models~\cite{Jaeckel:2010ni,Essig:2013lka,Murata_2015}, and may not have been detected either because they appear at weaker couplings or shorter distances than those at which existing searches have been sensitive~\cite{Murata_2015}.  
While here we focus on searches for BSM forces, similar techniques may also enable new measurements of extremely weak forces within occurring in the SM.  For example, the Casimir force~\cite{Lamoreaux:2012} is discussed below since it poses a significant background to some of the proposed searches~\cite{Geraci:2010,Onofrio:2006mq}, although its measurement with precision $\lesssim 10$\% remains a significant experimental challenge, for which optically levitated objects may provide new approaches.  In addition, measurements of similar effects such as the Casimir torque maybe be enabled by optically levitated objects~\cite{Xu:2017bpr,Ahn:2020}.  In the following sections we describe the motivation for searches for BSM forces, and the sensitivity that might be achieved by levitated systems. 

\subsubsection{Newton's law}
\label{sec:newton}
Tests of the inverse-square law (ISL) of gravity have been pursued at increasing levels of precision since the original torsion balance experiments performed by Cavendish~\cite{Murata_2015}.  This series of experiments has culminated in the extreme precision of modern torsion balances, including those operated by the E{\"o}t-Wash collaboration~\cite{EotWash2020} and at HUST~\cite{HUST:2020}.  Recent results from E{\"o}t-Wash now detect gravitational interactions down to a length scale of $\lambda = 52\ \mu$m~\cite{Kapner:2007,EotWash2020}.  While possible deviations below this length scale have much weaker constraints, at longer length scales any modifications to gravity or new forces can only provide small corrections to the ISL.  

There are substantial challenges to measure gravity-strength interactions at distances below 50~$\mu$m.  First, the total source mass that can be used scales as $M \propto \rho \lambda^3$ for a mass of density $\rho$.  As is done in torsion balance experiments, large area flat source masses can be used instead to limit the scaling to $M \propto \rho \lambda$, but background forces, which often scale with surface area, quickly become dominant.  In particular, patches of electric charge or other stray electric and magnetic forces can easily overwhelm the tiny forces of interest for $\lambda \lesssim 10\ \mu$m and below~\cite{Garrett:2020,Garrett:2015bde}.  At these length scales, fifth forces or modifications to gravity that are much larger than expected from the ISL are consistent with existing experiments, the most sensitive of which make use of micro-mechanical oscillators~\cite{Deca:2016} or cantilevers~\cite{Geraci:2008}.

\begin{figure}[t]
    \centering
    \includegraphics[width=0.6\columnwidth]{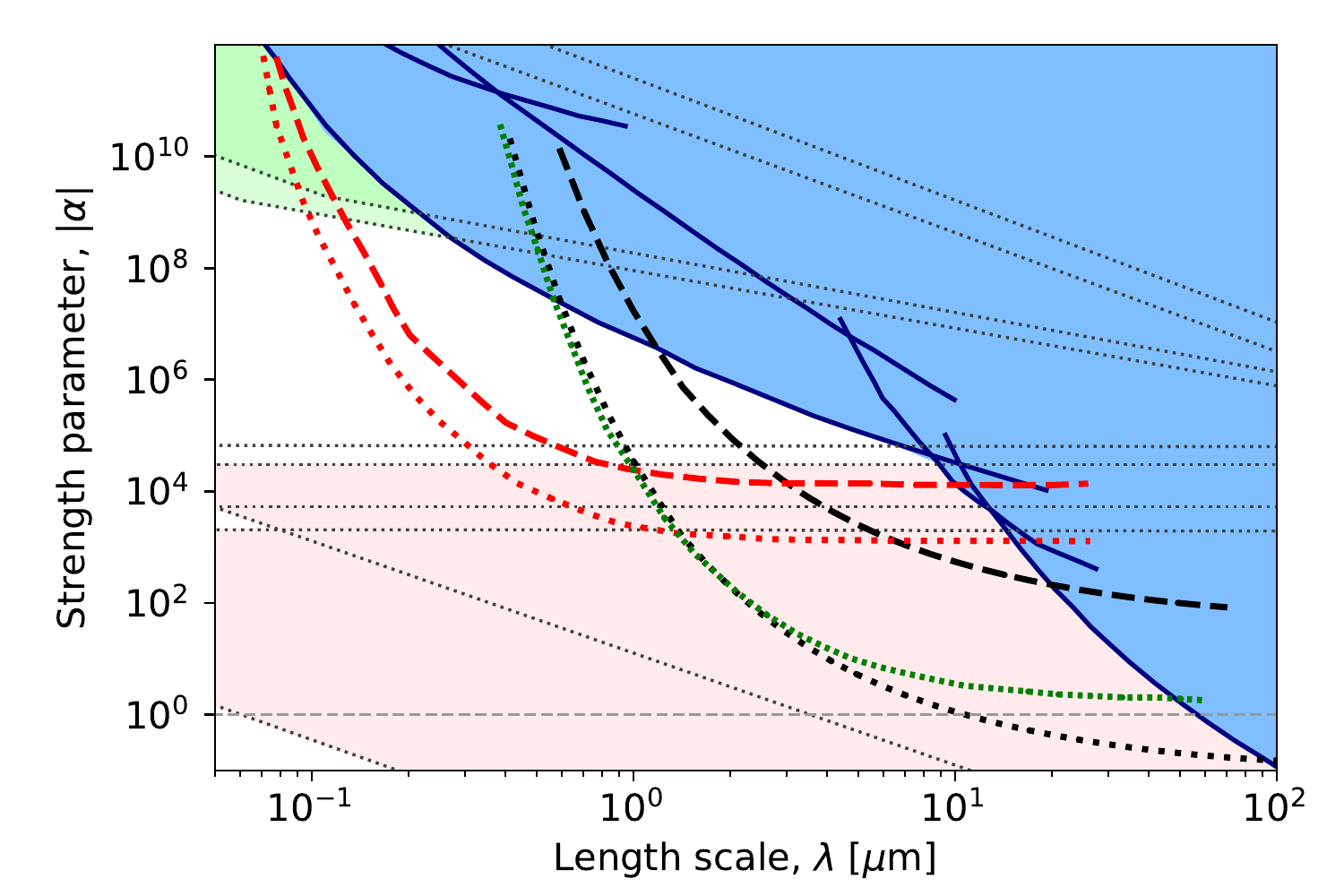}
    \caption{ Background free sensitivity projections to deviations from Newton's law for example optically levitated masses.  Existing limits are denoted by the blue region~\cite{Murata_2015,Deca:2016,EotWash2020}, with examples of specific theoretical predictions for such deviations in a selection of models of BSM physics (e.g., light moduli in string theories or supersymmetric extensions to the SM) denoted in red and green~\cite{PhysRevD.68.124021,Geraci:2008}. The black dashed line shows the projected sensitivity for a 20~$\mu$m diameter sphere at the best currently demonstrated sensitivity for a sphere of this size~\cite{acceleration2020} for a $10^5$~s integration, assuming no backgrounds. The black dotted line shows the corresponding sensitivity at the SQL.  The red dashed/dotted lines show the sensitivity possible for a nanosphere with diameter of 300~nm~\cite{Geraci:2010} limited by thermal noise at $10^{-8}$ torr (dashed) and ultimately by quantum photon-recoil heating (dotted).  The green dotted line shows the projected sensitivity for a light-pulse matter-wave interferometer employing 13~nm diameter spheres that are free-falling near a source mass surface ~\cite{Geraci:matter_wave}. }
    \label{fig:gravity_sens}
\end{figure}

Optically levitated objects may have advantages for tests of Newton's law in the range of length scales between $100\ \mathrm{nm} \lesssim \lambda \lesssim 50\ \mu$m.  In particular, the net charge of a nanosphere or microsphere can be precisely controlled, as described above mitigating backgrounds due to stray charge.  Higher order multiple moments of the charge distribution within the sphere (or ``patch potentials'' on its surface~\cite{Garrett:2015bde,kim2008anomalies}) can be controlled by rapidly rotating the sphere about the optical axis of the trap, at frequencies much above the measurement frequency~\cite{arita2016rotational,Rotation_paper_2018,Tongcang_rotation:2018,Rene:2018}.  Optical trapping provides substantial flexibility for positioning masses around the trapped sphere.  Long working distances~\cite{Moore:2014,Acceleration_2017} can enable shielding electrodes and attractors to be inserted into the region next to the trap~\cite{Kawasaki:2020oyl,Moore:2018spie}, which has been demonstrated at few $\mu$m spacing~\cite{Rider:2016}. Alternatively, sub-wavelength spheres can be trapped at separations of $a = \lambda/4$ from a reflective surface~\cite{Geraci:2010} (note here that $\lambda$ refers to the wavelength of light rather than characteristic range of a new force).  Finally, the acceleration sensitivities already achieved by $\mu$m scale levitated masses would be sufficient to surpass currently existing sensitivities, as shown in Fig.~\ref{fig:gravity_sens}.  Substantial improvement beyond the best existing tests could be reached if backgrounds and noise could be improved to allow operation at the SQL. 

\begin{figure}[t]
    \centering
    \includegraphics[width=\columnwidth]{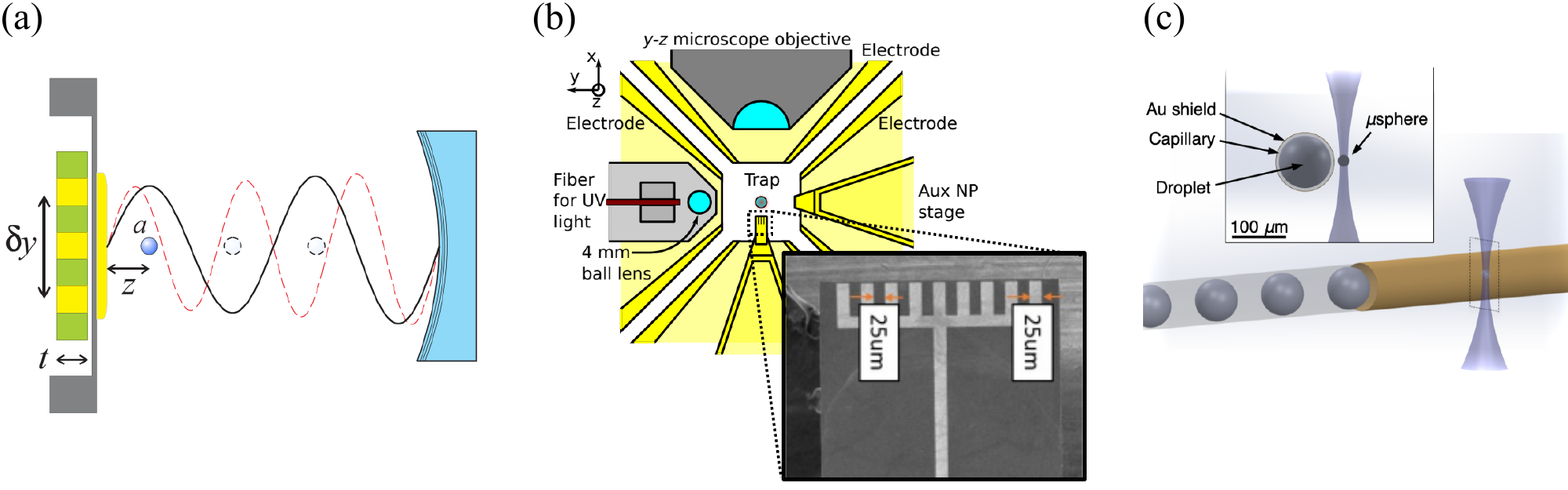}
    \caption{ Attractor designs for short-range gravity measurements using levitated systems proposed to-date.  (a) A subwavelength nanosphere is trapped at the antinode of a standing wave potential, several hundred nm from a gold coated membrane shield.  A patterned mass with different density materials (green and yellow) is oscillated behind the shield. Figure reproduced from~\cite{Geraci:2010}. (b) A linear array of Au and Si fingers, each 25~$\mu$m wide, is oscillated at several $\mu$m distance from a 5~$\mu$m diameter sphere trapped in the center of the surrounding shielding electrodes~\cite{Wang:2017}.  An additional microfabricated shield between the moving fingers and sphere can be positioned on an independent stage~\cite{Kawasaki:2020oyl}.  Figure reproduced from~\cite{Wang:2017,Kawasaki:2020oyl}.  (c) A Au-coated microfluidic capillary is positioned at several $\mu$m distance from a $\sim$10-20~$\mu$m diameter sphere, and alternating droplets of a dense fluid (e.g. a polytungstate salt solution, $\rho \approx 3$~g/cm$^3$) and carrier fluid (e.g. mineral oil, $\rho \approx 0.8$~g/cm$^3$) flow through the capillary at $\sim$50~Hz~\cite{Moore:2018spie}.}
    \label{fig:attractors}
\end{figure}

The primary challenge to implement such experiments is to position optically trapped objects near a mass attractor without introducing backgrounds that overwhelm the gravity-like force being searched for.  Fig.~\ref{fig:attractors} shows several current proposals for implementing such attractors. Several sources of backgrounds must be mitigated by any successful attractor design, including those arising from vibrations, scattered light, the Casimir force, and patch potentials.  Each of the attractor designs in Fig.~\ref{fig:attractors} relies on modulation of the mass density in the vicinity of the sphere (behind a stationary, electrically grounded shield) to avoid backgrounds arising from static forces.  These include the Casimir force between the sphere and the conductive surface of the attractor, which can exceed the gravity-like forces of interest by several orders of magnitude at $\lesssim\mu$m distances~\cite{Geraci:2010}.  Such shields have been successfully demonstrated to mitigate such backgrounds in larger scale apparatuses employing torsion balances~\cite{EotWash2020,Schmole:2016mde,Westphal:2020okx}.  However, motion of the attractor mass (required to produce such modulation) can induce both vibrations or scattered light backgrounds that lead to either real or apparent forces on the sphere at the frequency of modulation of the mass~\cite{Kawasaki:2020oyl}.  While such backgrounds are not fundamental in nature, they present a key technological challenge in implementations of optimized attractors~\cite{Geraci:2010,Kawasaki:2020oyl,Moore:2018spie}.  Finally, at $\lesssim\mu$m separations, significant backgrounds from electric forces arise due to patch potentials~\cite{Garrett:2015bde}, i.e. localized potential differences in microscopic patches of the conductive attractor surface that are difficult to eliminate. Both the control of the charge state of the sphere (as described above) and the use of a stationary shield can be used to reduce the effect of patch potential backgrounds at the measurement frequency. 

Mitigating the backgrounds described above is a formidable experimental challenge, which has not yet been demonstrated at the required sensitivity.  It remains to be determined to what level such background reduction is technically feasible, and this will set the ultimate reach of optically levitated systems to such forces.  However, if such backgrounds could be controlled below the current (and future) sensitivity of levitated optomechanical systems the resulting projected sensitivity is shown in Fig.~\ref{fig:gravity_sens}.  Here, deviations from the ISL are parameterized with a Yukawa form~\cite{Murata_2015}:
\begin{equation}
    V(r) = \frac{G_N m M}{r}(1 + \alpha e^{-r/\lambda})
\end{equation}
where $G_N$ is Newton's constant, $m$ and $M$ are the sphere and attractor masses (assumed to be point-like), $r$ is their separation, $\alpha$ parameterizes the strength of the new force, and $\lambda$ parameterizes its characteristic range.  If the deviation from the ISL is mediated by a new force carrier with mass $m_\phi$, then typically $\lambda = \hbar /(m_\phi c$).  While other forms for the deviation from the ISL are possible~\cite{Murata_2015}, Fig.~\ref{fig:gravity_sens} shows the constraints on the Yukawa form to enable a consistent comparison between multiple experiments.

Matter wave interferometry with nanospheres can also potentially be used to measure short-range surface forces such as the Casimir effect and tests of ISL violation. 
By cooling the center-of-mass motion of the nanosphere sufficiently, it should be possible to obtain a cold, point-like source for a matter wave interferometer \cite{Ulbricht:2014,Geraci:matter_wave}.   Atom-based interferometers have demonstrated remarkable acceleration sensitivity (below $10^{-11}$ $g/\sqrt{\rm{Hz}}$), where $g$ is the gravitational acceleration at the surface of the earth \cite{Kovachy:2015}. Nanosphere interferometers have the potential to achieve similar sensitivity but with a key advantage---the large mass of the nanosphere (compared with an atom) makes its wave-packet expand slowly enough to allow localized sensing at the micron scale \cite{Geraci:matter_wave}. This permits the particle to remain localized at micron distance from a surface over 1 second time scales. On the other hand, free-fall atom interferometers involve significantly more spread of the wave packet \cite{Kovachy:2015}, making them impractical for operation at micron separation from a surface. 
Using this approach, in the longer term future, it may be possible to probe new forces with $|\alpha|=10^4$ to $|\alpha|=1$ times the strength of gravity at ranges of $1$ to $10$ microns, respectively, potentially surpassing the non-interferometric methods at the shortest distances, while using a complementary, fully quantum approach, as shown in Fig. \ref{fig:gravity_sens}. Measurements of the Casimir effect are also possible with this method \cite{Geraci:matter_wave}.

\subsubsection{Coulomb's law}
Similar to the searches for new forces that couple to mass described in Sec.~\ref{sec:newton}, optically levitated systems may also be used to search for forces that couple to charge.  In particular, this includes searching for deviations from Coulomb's law, which may be motivated by the presence of new forces under which dark matter could be charged~\cite{Jaeckel:2009dh,Jaeckel:2010}.  Recent theoretical work has motivated study of a broad class of ``hidden sector'' dark matter models~\cite{Essig:2013lka}, in which dark matter may consist of multiple constituents charged under new forces that do not directly interact with SM particles.  In these models, such new forces can weakly mix with electromagnetism, resulting in new Coulomb-like interactions~\cite{Jaeckel:2010ni}.  A prototypical example is a new U(1) gauge group in the hidden sector~\cite{HOLDOM:1986,Jaeckel:2010ni}, for which a dark photon can kinetically mix with the standard model photon leading to a Yukawa potential (for two electrons of charge $e$):
\begin{equation}
    V(r) = \frac{\alpha_{EM}}{r}(1 + \chi^2 e^{-r/\lambda})
    \label{eq:coul}
\end{equation}
where $\alpha_{EM} = e^2/(4\pi\epsilon_0 \hbar c)$ is the fine structure constant, $\chi \ll 1$ is the coupling, and the length scale, $\lambda = \hbar/(m_\chi c)$ for a mediator  (e.g. a dark photon) with mass $m_\chi$.

Optically levitated objects have similar advantages for tests of Coulomb's law as for testing Newton's law---i.e., their mass is confined at $\mu$m or sub-$\mu$m length scales, allowing forces with mediator masses $m_\chi \lesssim 1$~eV to be probed.  In addition, the charge state of the sphere can be precisely controlled, and changed in polarity.  Net charges as large as $10^4\ e$ have been demonstrated for 10~$\mu$m diameter spheres, and larger charges may be possible~\cite{acceleration2020,Millen:2018}.  In addition, polarizing the sphere in an external electric field can allow induced dipole moments as large as $\sim$1~$e$\,cm in a 1~kV/mm electric field for a 10~$\mu$m diameter sphere. Such dipole moments can directly couple to electric field gradients, which may be advantageous for searching for corrections to Coulomb's law of the form in Eq.~\ref{eq:coul}, since for length scales $\sim \hbar/(m_\phi c)$, the gradient in the electric field over the diameter of the sphere can be a substantial fraction of its overall magnitude.  However, when compared to trapped ions (where electric field sensors have been demonstrated with sensitivity of $<1$~nV/m~\cite{Gilmore:2017cbw}), the charge-to-mass ratio of the sphere is roughly 9--10 orders-of-magnitude smaller, substantially reducing the expected acceleration from a deviation arising from Coulomb's law.  Thus, the tradeoffs between single trapped ions, and charged (or polarized) nano- or micro-scale objects primarily arise from the ability to mitigate backgrounds, which may provide limits in practical applications rather than the sensor's intrinsic sensitivity.

\begin{figure}[t]
    \centering
    \includegraphics[width=0.6\columnwidth]{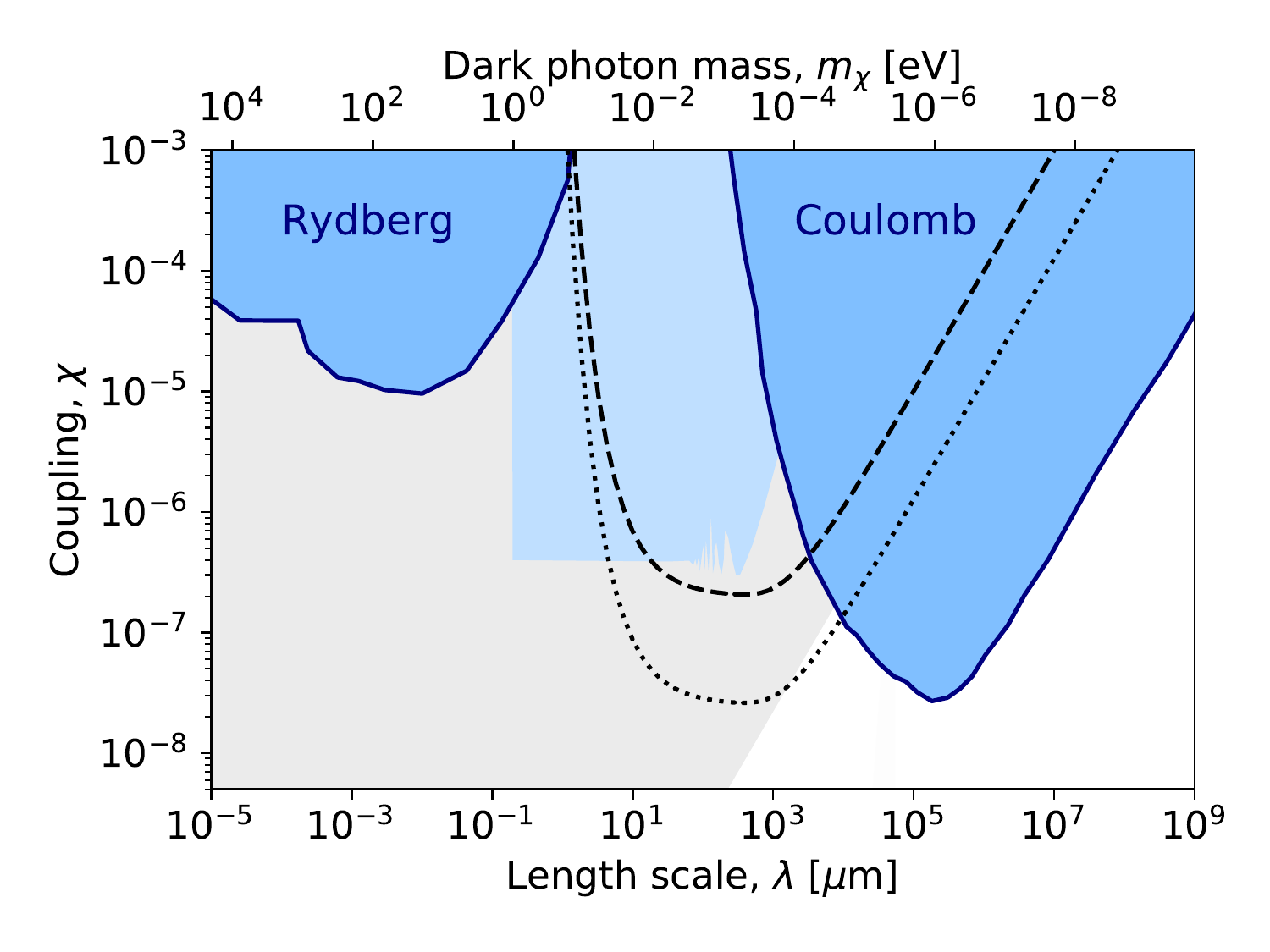}
    \caption{Background free sensitivity projections to deviations from Coulomb's law for a 20~$\mu$m diameter sphere at the best current demonstrated sensitivity~\cite{acceleration2020} (dashed) and the SQL (dotted).  Existing laboratory limits from Coulomb's law tests are denoted by the dark blue regions~\cite{Essig:2013lka}. The upper axis indicates the corresponding mass of a dark photon that could mediate the new force, with existing laboratory limits on dark photons from light shining through wall experiments denoted in light blue~\cite{EHRET2010149}. The gray region indicates astrophysical bounds on dark photons versus mass from stellar cooling constraints~\cite{Essig:2013lka}.  }
    \label{fig:coulomb_sens}
\end{figure}

Fig.~\ref{fig:coulomb_sens} shows the projected sensitivity of a 20~$\mu$m diameter optically levitated microsphere to deviations from Coulomb's law of the type parameterized in Eq.~\ref{eq:coul}, for a microsphere positioned just outside a fully shielded parallel plate capacitor with plate spacing of 2~mm and a separation between the sphere and ground electrode of 25~$\mu$m.  Rather than modulating a mass density as the source, only a modulating AC electric field between the plates of the parallel plate capacitor is required. In the case that Coulomb's law holds exactly, the field outside the capacitor must vanish.  However, for corrections to Coulomb's law of the type described in Eq.~\ref{eq:coul}, a small leakage of the field (proportional to $\chi^2$) outside the plates into the region of the sensor is expected. Such a search could provide the most sensitive laboratory test of Coulomb's law at length scales between $1\ \mu\mathrm{m} < \lambda < 1\ \mathrm{mm}$.  However, if such deviations arose from a model with a dark photon, astrophysical limits from stellar cooling would also rule out much of the accessible parameter space with microspheres~\cite{Essig:2013lka}.  Whether such sensitivities can be reached largely depends on the background level achievable.  Technical sources of backgrounds including vibrations of the electrodes, scattered light, stray electric fields, and electrical pickup would all need to be controlled to reach the projected sensitivities.

\subsubsection{Matter neutrality}
One of the first applications of optically levitated systems to searches for BSM physics was to test the electric neutrality of nanogram-scale masses with sensitivity to single fractional charges $\gtrsim 10^{-4}\ e$~\cite{Moore:2014}.  Such tests can be sensitive either to particles with tiny fractional charges bound in matter (i.e., ``millicharged'' particles with charge $\ll 1\ e$), or a difference in the sum of the proton, neutron, and electron charge from zero. 

Precision measurements over the past century have demonstrated that the electric charges of the proton and electron are equal and opposite to within at least one part in $10^{21}$~\cite{Unnikrishnan_2004}. The experimental observation that these fundamental building blocks of matter carry charges that must be nearly exactly quantized in units of $\pm$1~$e$ may provide hints of some deeper underlying symmetry between quarks and leptons that is not yet incorporated in the SM~\cite{PDG2018}. If such an underlying symmetry does not exist in nature, it may still be possible to obtain exact charge quantization through requirements on gauge anomaly cancellation, but such relations are typically sensitive to the full content of particles existing in nature and thus may provide a new probe for BSM physics~\cite{Foot:1993yq}.  

In addition, fractionally charged particles have been searched for in many experiments since the discovery of quantized charges (in particular, searches for free quarks with charge $e/3$).  Modern motivations for searching for fractional charges $\ll 1e$ arise in hidden sector dark matter models~\cite{Essig:2013lka}.  If a new force in the hidden sector weakly mixes with SM electromagnetism, then dark matter particles can obtain tiny effective charges~\cite{Essig:2013lka,Jaeckel:2010ni}.  Such particles could be produced in the early universe and form bound states with atoms.  If such particles are $\gtrsim 10$~GeV in mass and have fractional charge, $q \gtrsim 10^{-3}\ e$, then eV-scale binding energies are possible in a Bohr-like ``atom,'' for which a millicharged DM particle and nucleus are bound.  Such massive particles may also have escaped detection at particle accelerators, where dedicated searches probe masses only $\lesssim 1$~GeV~\cite{Ball:2020dnx, Prinz:1998ua}.

\begin{figure}[t]
    \centering
    \includegraphics[width=\columnwidth]{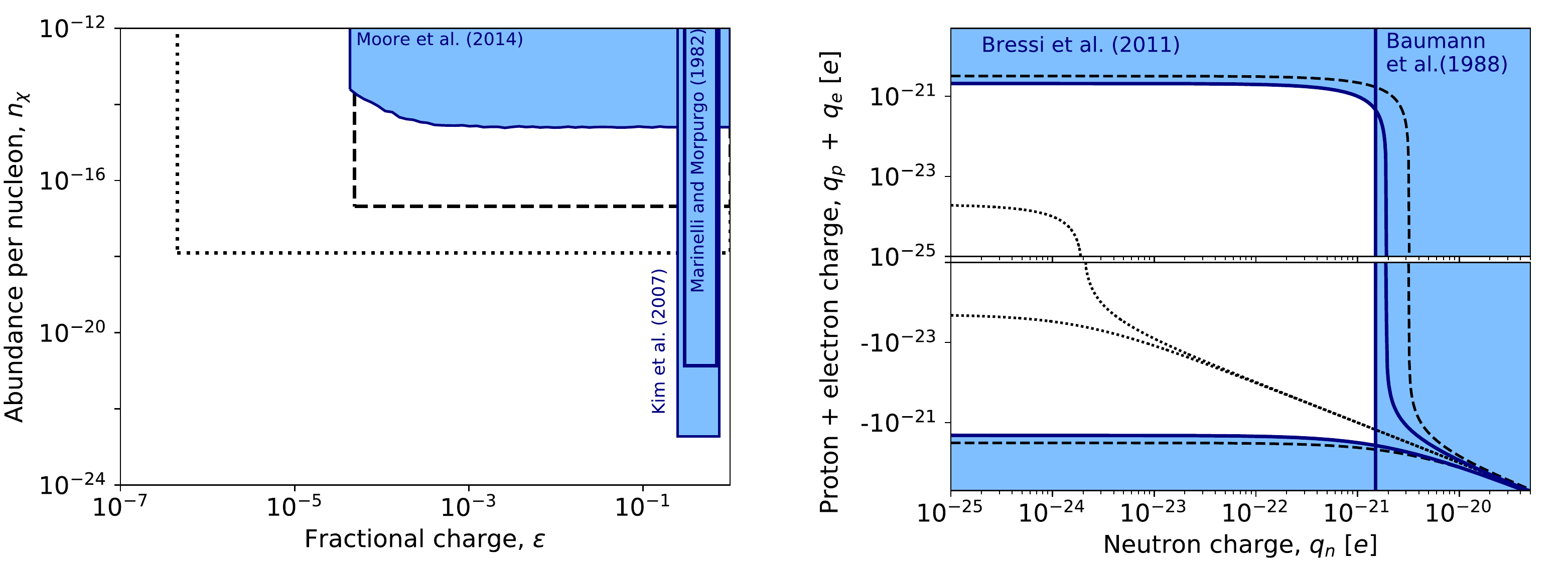}
    \caption{ Background free sensitivities for millicharged particles (left) and non-neutrality of matter (right) for a 20~$\mu$m diameter sphere at the best current demonstrated sensitivity~\cite{acceleration2020} (dashed) and the SQL (dotted). In both cases, existing limits are denoted by the blue regions~\cite{Moore:2014,PhysRevLett.99.161804,Morpurgo:1984,Bressi:2011,Baumann:1988}.}
    \label{fig:matter_neut}
\end{figure}

Levitated objects have a long history in testing the neutrality of matter and searching for fractionally charged particles.  While the most sensitive current tests of the neutrality of matter employ neutrons~\cite{Baumann:1988} or macroscopic acoustic resonators~\cite{Bressi:2011}, magnetically levitated spheres used for free quark searches reached comparable sensitivities~\cite{Morpurgo:1984}. Smaller particles typically provide more sensitivity for small fractional charges bound in matter, and recent modernized implementations of the Millikan oil drop experiment have searched for fractional charges $>e/3$ in nearly $\sim 5\times 10^7$ droplets, giving a total mass of 0.26~g~\cite{PhysRevLett.99.161804}.  Ashkin first proposed the use of optically levitated spheres to perform a modern, ultra-sensitive version of the Millikan experiment in 1980~\cite{Ashkin1980}, and as described above, results of such an experiment were first reported in 2014~\cite{Moore:2014}.  Future extensions of these techniques may continue to substantially improve sensitivity to tiny fractional charges or non-neutrality of matter, as shown in Fig.~\ref{fig:matter_neut}. 

A key advantage of levitated systems for tests of the neutrality of the objects is that their charge state can be precisely controlled and measured. The primary challenge to practical implementations result from the presence of permanent dipole moments within the spheres~\cite{Rider:2016,Moore:2014}.  However, these dipole moments can be controlled through rapid rotation of the sphere, either optically~\cite{Rotation_paper_2018,arita2016rotational,Tongcang_rotation:2018,Rene:2018} or electrically~\cite{Rider:2019}.  The sensitivity that is in principle achievable is significant---for a 10~$\mu$m diameter sphere with already demonstrated force sensitivity of 1~aN/$\sqrt{Hz}$ and acceleration sensitivity of 100~n$g/\mathrm{\sqrt{Hz}}$ in a 1~kV/mm field, then a fractional charge sensitivity of $\epsilon \approx 6\times10^{-8}\ e$ could be reached in a $10^4$~s integration, assuming no other backgrounds.  This would correspond to a sensitivity to the sum of the proton, neutron, and electron charges of $|q_p + q_n + q_e| \approx 10^{-22}$, which is more than an order of magnitude beyond the best previous searches.  Longer integrations or lower noise could in principle allow even more sensitive tests to be performed. For example, sensitivity at the $10^{-28}\ e$ level could be achieved, in principle, assuming a background-free measurement of a 50~$\mu$m diameter sphere in a 10~kV/mm electric field for an integration of $10^5$~s at the SQL.  However, controlling backgrounds at the level of these sensitivities would require substantial advances beyond the state-of-the-art. Such sensitivity would be comparable to proposals involving free-fall atom interferometers, which have also suggested the possibility to test $|q_p + q_n + q_e|$ at the $10^{-28}\ e$ level \cite{atom2008}.

Figure~\ref{fig:matter_neut} shows the projected sensitivity of optically levitated spheres to fractional charges and the neutrality matter, although at more modest sensitivities than those described just above.  In particular, it is assumed that backgrounds can be controlled sufficiently that a $\sim$1~s integration is possible at current (and future) sensitivities.  This more conservative projection is based on recent experience in the practical difficulty of mitigating backgrounds in such measurements~\cite{Moore:2014,Rider:2016}. Indeed, even if the dipole (and higher order multiple) backgrounds can be suppressed at the required level, technical sources of backgrounds arising from vibrations (large electrical forces will be present on the electrodes used to produce the fields and HV feedthroughs, although typically at twice the drive frequency) or electrical pickup can present practical limits.  However, if such technical backgrounds can be overcome, optically levitated nanogram masses have the advantage that their charge state is precisely known, avoiding the dominant irreducible background that prevents progress using macroscopic resonators~\cite{Bressi:2011}.  Further work to control technical backgrounds present for optically levitated spheres could thus allow sensitivity beyond even that projected in Fig.~\ref{fig:matter_neut} for millicharged particles and non-neutrality of matter.

\subsection{Searches for weak astrophysical signals}

The extreme force sensitivity made possible by optical levitation lends itself to the search for weak astrophysical signals, including feeble strain signals from gravitational waves or impulses from passing dark matter.

\subsubsection{High-frequency gravitational waves}

\begin{figure}[t]
    \centering
    \includegraphics[width=0.7\columnwidth]{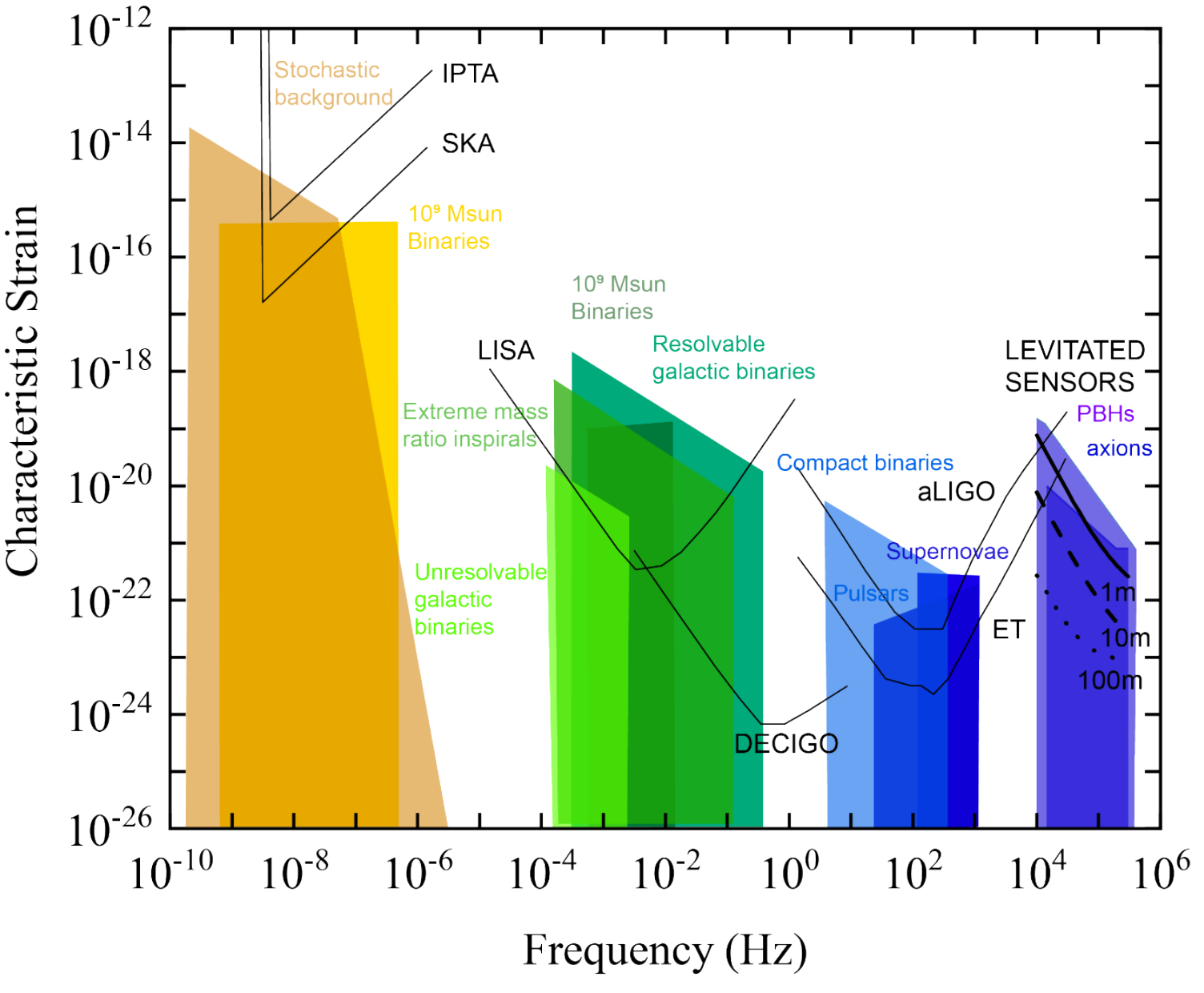}
    \caption{Adapted from Ref. \cite{GWMoore_2014}. Frequency landscape of gravitational wave searches for various technologies. }
    \label{fig:GW_landscape}
\end{figure}

One of the most interesting sources of gravitational waves in the high-frequency regime arises from physics Beyond the Standard Model. The QCD axion is a well-motivated dark matter candidate that naturally solves the strong CP problem in strong interactions and explains the smallness of the neutron's electric dipole moment \cite{axion1,axion2,PTViolation,Moody:1984ba}. The mass of a QCD axion, $m_a$ is related to its decay constant, $f_a$, as $m_a = 5.7\ \mathrm{meV}\ (10^9\ \mathrm{GeV}/f_a)$~\cite{PDG2018}.  For a QCD axion with decay constant $f_a \sim 10^{16}$ GeV (at the Grand-Unified-Theory [GUT] energy scale), the Compton wavelength of the axion matches the size of stellar mass BHs and allows for the axion to bind with the BH ``nucleus,'' forming a gravitational atom in the sky. A cloud of axions grows exponentially around the BH, extracting energy and angular momentum from the BH \cite{stringaxiverse,stringaxiverse2}. Axions in this cloud produce gravitational radiation through annihilations of axions into gravitons. For annihilations, the frequency of the produced GWs is given by twice the mass of the axion: for example, $f=145\ \mathrm{kHz}$ when $f_a$ is around the GUT scale, which lies in the optimal sensitivity range for optically levitated sensors. The signal is coherent, monochromatic, long-lived, and thus completely different from all ordinary astrophysical sources \cite{minamasha2014}. The fraction of the BH mass the axion cloud carries can be as high as ${10}^{-3}$ \cite{stringaxiverse2}, leading to strain signals detectable within the sensitivity band of optically levitated sensors \cite{GWprl}. Fig. \ref{fig:GW_landscape} shows the anticipated search reach of levitated sensor technology along with existing and planned gravitational wave observatories and detection methods. The reach of the 1-meter LSD (Levitated sensor detector) \cite{GWlev2020} being built currently is indicated as well as for future length upgrades.
In recent work, an alternative matter-wave interference approach has been proposed to search for GWs in the frequency range below 10 Hz \cite{SougatoGW}, using macroscopic meter-scale quantum superpositions of levitated nanodiamonds. 

\subsubsection{Dark matter}
\label{sec:DM}
The nature of dark matter (and its detection in terrestrial experiments) remains among the most important open questions in fundamental physics.  Most existing laboratory searches for dark matter are optimized to detect either individual nuclear or electron recoils in a massive target~\cite{Schumann:2019eaa}, or the production of a photon (or phonon) from dark matter conversion in a resonant cavity or structure~\cite{Graham:2015ouw}.  However, for long-range interactions between dark matter and normal matter (i.e., those with an effective range much larger than atomic spacings), recoils of sub-atomic particles may not provide the optimal search strategy.  In these cases, collective modes of many atoms can typically be excited with larger energy or momentum transfers than single electrons or nuclei~\cite{Knapen:2017ekk,Coskuner:2019,Griffin:2018bjn,Monteiro:2020wcb}.  

In fact, the only interaction that dark matter is known to have is long-range in nature (i.e., its gravitational interaction, mediated by a massless force carrier).  While detection of dark matter via gravity alone may be possible in principle~\cite{carney:2019_1,Carney:2020xol,Hall:2018,AkioDMpaper}, it would require substantial improvements in optomechanical sensors beyond the current state-of-the-art, and likely larger masses than could be optically levitated.  Thus, optically levitated systems are most useful in the case that DM couples to normal matter via a long-range force that could exceed gravity in strength.  This sort of ``fifth force'' coupling to mass is exactly of the type described in Sec.~\ref{sec:newton}.  If DM also coupled to such a force, it could lead to new detection strategies~\cite{Carney:2020xol,Cheng:2019vwy,Lee:2020dcd,Bateman:2014lia,Riedel:2016acj} for relic DM particles.

\begin{figure}[t]
    \centering
    \includegraphics[width=0.6\columnwidth]{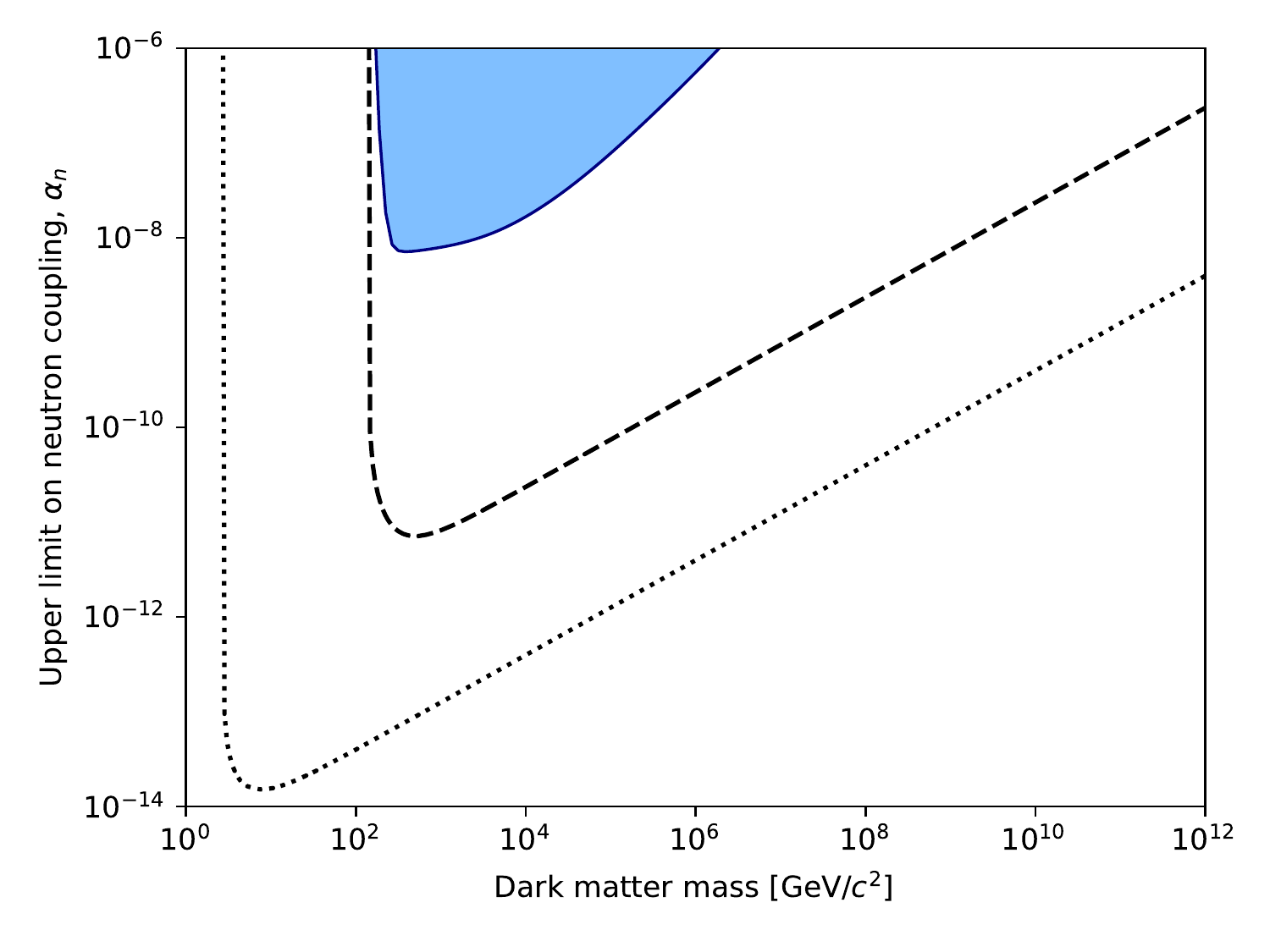}
    \caption{Background free sensitivity projections to a dark matter candidate interacting with neutrons through a long-range force, assuming a massless mediator.  Existing constraints from an initial demonstration of such a search (with a few ng-days exposure) are shown by the blue filled region~\cite{Monteiro:2020wcb}.  Substantial improvement over the initial search is possible with longer exposures and larger arrays of spheres. Projections for the sensitivity of a 100~element array of spheres with 1~yr exposure at the current sensitivity (dashed) and at the SQL (dotted) are also shown.   }
    \label{fig:dm_sens}
\end{figure}

A recent search has been performed for composite dark matter particles scattering from an optically levitated nanogram mass, cooled to an effective temperature $\sim$200~$\mu$K~\cite{Monteiro:2020wcb}. This search placed limits on the interaction strength between DM and neutrons, $\alpha_n \leq 1.2 \times 10^{-7}$ for a new long-range force with mediator mass, $m_\phi \leq 0.1$~eV for dark matter masses in the range 1--10~TeV.  Here, $\alpha_n$ denotes any generic coupling for which the potential between dark matter and a point mass containing $N$ neutrons is:
\begin{equation}
\label{yukawa}
V(r) =   \frac{ \alpha_n N}{r}e^{-m_\phi r (c/\hbar)}
\end{equation}
While these constraints are presented as a generic coupling, they can be compared to specific model-dependent constraints from existing detectors after assuming a microscopic model for dark matter~\cite{Coskuner:2019,Monteiro:2020wcb}.  In general, these searches can probe models inaccessible to large WIMP detectors or other searches aiming to detect single nuclear or electron recoils for dark matter particles that collectively scatter from multiple targets, and transfer sufficiently small energy to normal matter that detectable recoils would not be produced in existing detectors~\cite{Monteiro:2020wcb}.
Since the scattering cross section scales as $\sim 1/q^4$, where $q$ is the momentum transferred to the sphere, future searches with lower detection thresholds (or consisting of arrays of many spheres) could further improve sensitivity to these models by many orders of magnitude~\cite{Carney_white_paper}, as shown in Fig.~\ref{fig:dm_sens}.  

Unlike previous applications to tests of force laws described in Sec.~\ref{sec:forces}, background-free sensitivity is likely to be possible in such searches due to their inherent directional sensitivity~\cite{Ahlen:2009ev}, although sources of background impulses such as vibrations and electrical pickup must still be mitigated. The key technical challenge to reaching large exposures is to construct large arrays of optically levitated sensors. One possible path to integrating such arrays would be to miniaturize the traps using photonic integrated circuits~\cite{2020arXiv200501948W}.  Alternatively, tweezer arrays using spatial light modulators or acousto-optics could be implemented following the successful technologies demonstrated for cold atoms~\cite{2016Sci...354.1021B,2016Sci...354.1024E}.

\subsection{Quantum foundations and tests of gravitational effects in quantum theory}

Quantum mechanics as currently formulated should apply equally well to the macroscopic and microscopic worlds. Notwithstanding decoherence through interactions with their environment, it contains no \textit{fundamental} mechanism for preventing macroscopic objects from existing in a coherent linear superposition of two different quantum states.  Since the early days of quantum mechanics, this fact has deeply bothered physicists, as it conflicts with the absence of macroscopic quantum superpositions in everyday life.  Moreover, the lack of a fundamental mechanism to prevent macroscopic superpositions is closely related to the measurement problem in quantum mechanics.  In order to produce agreement with experimental results, some theorists have formulated an ad hoc postulate of “measurement induced collapse” to describe the outcomes of measurements of a microscopic system in a superposition of two quantum states, using a macroscopic device.  By contrast, at all other times, the evolution a quantum system is described by unitary evolution under the Schr$\ddot{{\rm{o}}}$dinger equation.  

\subsubsection{Quantum collapse models}

The dissonance between how quantum mechanics describes ordinary unitary evolution versus the measurement process is widely seen as unsatisfactory \cite{BassiRMP}.  As a potential resolution of these issues, models have been proposed that explain the absence of observations of macroscopic quantum superpositions and eliminate the need for the “measurement induced collapse” postulate by providing a formal mechanism for the collapse of superpositions of sufficiently massive objects.  In several of these models, the collapse is gravity-induced \cite{Penrose1996,Bassi:2017,BassiRMP,Diosi1989}.  Experimentally testing such collapse models probes the interface between quantum mechanics and gravity, and a confirmation of gravity-induced collapse would be a revolutionary advance in our understanding of quantum theory and reveal an interplay between gravity and quantum mechanics that would need to be considered in developing any theory of quantum gravity. Improving tests of collapse models beyond previous work can be achieved by creating quantum superpositions involving more massive objects and longer coherence times.  A promising approach to achieve this is interferometry with levitated nanoparticles \cite{Oriol2011}. A figure of merit describing how macroscopic a quantum superposition is or its  ``macroscopicity'' has been defined for objects ranging from individual atoms to larger scale mechanical oscillators \cite{BassiRMP}.   Optically levitated systems represent a method to test spontaneous localization models at intermediate scales \cite{BassiRMP,Bassi:2017}. The current bounds are shown in Ref. \cite{Carlesso:19}.
We also note that collapse models also can lead to anomalous heating effects in both clamped and levitated mechanical oscillators, and can also be searched for with non-interferometric methods \cite{BassiRMP, Carlesso:19, Zheng2020,Vinante2020,VinantePRL2020}. A recent underground test of the Diose-Penrose model has also been performed by searching for radiation expected from the random motion of charged particles \cite{Donadi2020}.

Any fundamental decoherence mechanism whether gravitational related or not would need to be distinguished from technical sources of decoherence. One challenging aspect of optically levitated systems stems from the fact that for optically trapped particles in vacuum, the center of mass temperature can be quite cold while the internal  temperature  can  be  significantly  above  room  temperature.   As  such  a  warm  object  is  diffracted from  a  light  grating and  placed  into  a  superposition state, blackbody emission from  the  surface  of  the  spheres  can  decohere  the  quantum  superposition \cite{Ulbricht:2014,Geraci:matter_wave}.  Optical refrigeration techniques or operation in a cryogenic environment with minimal laser heating may be routes for improving the blackbody emission limited lifetime of such macroscopic quantum superpositions \cite{Rahman2017}. Magnetic levitation of superconducting particles has also been proposed as a method for obtaining larger quantum superpositions without the effects of laser heating \cite{Oriolracetrack}. 

\subsubsection{The role of gravity in entanglement}
While probing physics at the Planck scale directly may be outside of the realm of possibility for humankind in the near future, examining the role that gravity plays in uniquely quantum phenomena such as entanglement can provide insight into the quantum nature of gravity~\cite{PhysRevLett.47.979,2013arXiv1311.4558K, Carlesso:2019}. Proposals have recently been presented for using macroscopic superpositions of massive nanoparticles to test whether the gravitational field can entangle the states of two masses \cite{Marletto:2017,Bose:2017,Carlesso:2017vrw}.  In these proposals, it is described how the gravitational interaction alone between two quantum superpositions can create entanglement between the states of a massive object. The state of embedded spins in the masses can be used as a witness to probe the entanglement as described in Ref. \cite{Bose:2017}. A significant challenge in these experiments will be to avoid other non-gravitational interactions due to surface forces and other external environmental perturbations. Given the overall rate of progress in the field, despite these technical challanges, levitated naoparticles provide a tantalizing possibility for novel investigations into the quantum nature of the gravitational interaction within the coming decade. 

\section{Conclusions}
Optically trapped sensors have already developed to the point where a large number of BSM searches for new physics are possible in the coming years using demonstrated techniques.  These searches have the potential to explore broad classes of models for new BSM physics that were previously inaccessible.  As the sensitivities for such systems continue to approach (and possibly eventually surpass) the SQL in the mass range between $\sim$1~fg to $\sim$1~ng, the potential of these searches for probing BSM physics will continue to improve.  Given the high sensitivities already achieved, the key challenge in the coming years for many applications is to mitigate technical sources of backgrounds that may overwhelm the tiny effects that such sensors aim to detect. Continued development of techniques to control the translational and rotational degrees of freedom of these sensors, as well as their charge distribution will be crucial to mitigating such backgrounds. If successful, such developments will enable a new tool at the precision frontier of particle physics for exploring some of the most difficult unanswered questions in fundamental physics.

\section*{Acknowledgments}
We would like to thank G. Afek and F. Monteiro (Yale) for helpful comments on the draft of this manuscript. DCM is supported in part by NSF grant PHY-1653232, the Heising-Simons Foundation, the Alfred P. Sloan Foundation, and ONR Grant N00014-18-1-2409. AG is supported in part by NSF grants PHY-1806686 and PHY-1806671, the Heising-Simons Foundation, the W.M. Keck Foundation, and ONR Grant N00014-18-1-2370.\\

\bibliographystyle{iopart-num}
\bibliography{references}{}

\providecommand{\newblock}{}
\begin{thebibliography}{100}
\expandafter\ifx\csname url\endcsname\relax
  \def\url#1{{\tt #1}}\fi
\expandafter\ifx\csname urlprefix\endcsname\relax\def\urlprefix{URL }\fi
\providecommand{\eprint}[2][]{\url{#2}}

\bibitem{PDG2018}
Tanabashi M {\em et~al.\/} (Particle Data Group) 2018 {\em Phys. Rev. D\/} {\bf
  98}(3) 030001

\bibitem{Odom_g-2}
Odom B, Hanneke D, D'Urso B and Gabrielse G 2006 {\em Phys. Rev. Lett.\/} {\bf
  97}(3) 030801

\bibitem{Safronova:2017xyt}
Safronova M, Budker D, DeMille D, Kimball D~F~J, Derevianko A and Clark C 2018
  {\em Rev. Mod. Phys.\/} {\bf 90} 025008 (\textit{Preprint}
  \eprint{1710.01833})

\bibitem{Erler:2019_EW_precision}
Erler J and Schott M 2019 {\em Prog. Part. Nucl. Phys.\/} {\bf 106} 68--119
  (\textit{Preprint} \eprint{1902.05142})

\bibitem{Jaeckel:2010ni}
Jaeckel J and Ringwald A 2010 {\em Ann. Rev. Nucl. Part. Sci.\/} {\bf 60}
  405--437 (\textit{Preprint} \eprint{1002.0329})

\bibitem{Kapner:2007}
Kapner D~J, Cook T~S, Adelberger E~G, Gundlach J~H, Heckel B~R, Hoyle C~D and
  Swanson H~E 2007 {\em Phys. Rev. Lett.\/} {\bf 98}(2) 021101

\bibitem{Yang2012}
Yang S~Q, Zhan B~F, Wang Q~L, Shao C~G, Tu L~C, Tan W~H and Luo J 2012 {\em
  Phys. Rev. Lett.\/} {\bf 108}(8) 081101

\bibitem{2011PhRvL.106d1801H}
{Hoedl} S~A, {Fleischer} F, {Adelberger} E~G and {Heckel} B~R 2011 {\em
  {Physical Review Letters}\/} {\bf 106} 041801

\bibitem{TheLIGOScientific:2014jea}
Aasi J {\em et~al.\/} (LIGO Scientific) 2015 {\em Class. Quant. Grav.\/} {\bf
  32}

\bibitem{Deca:2016}
Chen Y~J, Tham W~K, Krause D~E, L\'opez D, Fischbach E and Decca R~S 2016 {\em
  Phys. Rev. Lett.\/} {\bf 116}(22) 221102

\bibitem{Geraci:2008}
Geraci A~A, Smullin S~J, Weld D~M, Chiaverini J and Kapitulnik A 2008 {\em
  Phys. Rev. D\/} {\bf 78}(2) 022002 (\textit{Preprint} \eprint{0802.2350})

\bibitem{GiudiceDimopoulos}
{Dimopoulos} S and {Giudice} G~F 1996 {\em Physics Letters B\/} {\bf 379}
  105--114 (\textit{Preprint} \eprint{hep-ph/9602350})

\bibitem{add}
Arkani-Hamed N, Dimopoulos S and Dvali G 1998 {\em Physics Letters B\/} {\bf
  429} 263 -- 272

\bibitem{adelberger2020}
Lee J~G, Adelberger E~G, Cook T~S, Fleischer S~M and Heckel B~R 2020 {\em Phys.
  Rev. Lett.\/} {\bf 124}(10) 101101

\bibitem{HUST:2020}
Tan W~H, Du A~B, Dong W~C, Yang S~Q, Shao C~G, Guan S~G, Wang Q~L, Zhan B~F,
  Luo P~S, Tu L~C and Luo J 2020 {\em Phys. Rev. Lett.\/} {\bf 124}(5) 051301

\bibitem{wagnerEPV}
Wagner T~A, Schlamminger S, Gundlach J~H and Adelberger E~G 2012 {\em Classical
  and Quantum Gravity\/} {\bf 29} 184002

\bibitem{LIGOfirst}
Abbott B~P {\em et~al.\/} (LIGO Scientific Collaboration and Virgo
  Collaboration) 2016 {\em Phys. Rev. Lett.\/} {\bf 116}(6) 061102

\bibitem{Ashkin:1971}
{Ashkin} A and {Dziedzic} J~M 1971 {\em Appl. Phys. Lett.\/} {\bf 19} 283--285

\bibitem{Ashkin:1976}
{Ashkin} A and {Dziedzic} J~M 1976 {\em Appl. Phys. Lett.\/} {\bf 28} 333

\bibitem{Ashkin:1977}
{Ashkin} A and {Dziedzic} J~M 1977 {\em Appl. Phys. Lett.\/} {\bf 30} 202

\bibitem{Millen:2020review}
Millen J, Monteiro T~S, Pettit R and Vamivakas A~N 2020 {\em Rept. Prog.
  Phys.\/} {\bf 83} 026401 (\textit{Preprint} \eprint{1907.08198})

\bibitem{Li:2011}
Li T, Kheifets S and Raizen M~G 2011 {\em Nat. Phys.\/} {\bf 7} 527--530
  (\textit{Preprint} \eprint{1101.1283})

\bibitem{Gieseler:2012}
Gieseler J, Deutsch B, Quidant R and Novotny L 2012 {\em Phys. Rev. Lett.\/}
  {\bf 109}(10) 103603 (\textit{Preprint} \eprint{1202.6435})

\bibitem{Moore:2014}
Moore D~C, Rider A~D and Gratta G 2014 {\em Phys. Rev. Lett.\/} {\bf 113}(25)
  251801 (\textit{Preprint} \eprint{1408.4396})

\bibitem{Ranjit:2015}
Ranjit G, Atherton D~P, Stutz J~H, Cunningham M and Geraci A~A 2015 {\em Phys.
  Rev. A\/} {\bf 91}(5) 051805 (\textit{Preprint} \eprint{1503.08799})

\bibitem{Pettit:2019}
{Pettit} R~M, {Ge} W, {Kumar} P, {Luntz-Martin} D~R, {Schultz} J~T, {Neukirch}
  L~P, {Bhattacharya} M and {Vamivakas} A~N 2019 {\em Nature Photonics\/} {\bf
  13} 402--405

\bibitem{Huizhu:18}
Li W, Li N, Shen Y, Fu Z, Su H and Hu H 2018 {\em Appl. Opt.\/} {\bf 57}
  823--828

\bibitem{Vovrosh:2017}
{Vovrosh} J, {Rashid} M, {Hempston} D, {Bateman} J, {Paternostro} M and
  {Ulbricht} H 2017 {\em J. Opt. Soc. Am. B\/} {\bf 34} 1421 (\textit{Preprint}
  \eprint{1603.02917})

\bibitem{Aspelmeyer:2019}
Deli{\'c} U, Reisenbauer M, Dare K, Grass D, Vuleti{\'c} V, Kiesel N and
  Aspelmeyer M 2020 {\em Science\/} {\bf 367} 892--895

\bibitem{DUrso:2018}
Slezak B~R, Lewandowski C~W, Hsu J~F and D'Urso B 2018 {\em New J. Phys.\/}
  {\bf 20} 063028

\bibitem{Ulbricht:2019}
Vinante A, Falferi P, Gasbarri G, Setter A, Timberlake C and Ulbricht H 2020
  {\em Phys. Rev. Applied\/} {\bf 13}(6) 064027

\bibitem{BrianD'Urso_2020}
Lewandowski C~W, Knowles T~D, Etienne Z~B and D'Urso B 2020  (\textit{Preprint}
  \eprint{2002.07585})

\bibitem{Gieseler:2020}
Gieseler J, Kabcenell A, Rosenfeld E, Schaefer J~D, Safira A, Schuetz M~J~A,
  Gonzalez-Ballestero C, Rusconi C~C, Romero-Isart O and Lukin M~D 2020 {\em
  Phys. Rev. Lett.\/} {\bf 124}(16) 163604

\bibitem{Dania:2020kzl}
Dania L, Bykov D~S, Knoll M, Mestres P and Northup T~E 2020  (\textit{Preprint}
  \eprint{2007.04434})

\bibitem{Bullier2020}
{Bullier} N~P, {Pontin} A and {Barker} P~F 2020 {\em J. Phys. D Appl. Phys.\/}
  {\bf 53} 175302 (\textit{Preprint} \eprint{1906.09580})

\bibitem{Millen:2018}
{Goldwater} D, {Stickler} B~A, {Martinetz} L, {Northup} T~E, {Hornberger} K and
  {Millen} J 2019 {\em Quant. Sci. Tech.\/} {\bf 4} 024003

\bibitem{Ranjit:2016}
Ranjit G, Cunningham M, Casey K and Geraci A~A 2016 {\em Phys. Rev. A\/} {\bf
  93}(5) 053801 (\textit{Preprint} \eprint{1603.02122})

\bibitem{Acceleration_2017}
Monteiro F, Ghosh S, Fine A~G and Moore D~C 2017 {\em Phys. Rev. A\/} {\bf
  96}(6) 063841

\bibitem{acceleration2020}
Monteiro F, Li W, Afek G, Li C~l, Mossman M and Moore D~C 2020 {\em Phys. Rev.
  A\/} {\bf 101}(5) 053835

\bibitem{hempston2017force}
Hempston D, Vovrosh J, Toro{\v{s}} M, Winstone G, Rashid M and Ulbricht H 2017
  {\em Applied Physics Letters\/} {\bf 111} 133111

\bibitem{Novotny_static:2018}
Hebestreit E, Frimmer M, Reimann R and Novotny L 2018 {\em Phys. Rev. Lett.\/}
  {\bf 121}(6) 063602

\bibitem{Hoang:2016}
Hoang T~M, Ma Y, Ahn J, Bang J, Robicheaux F, Yin Z~Q and Li T 2016 {\em Phys.
  Rev. Lett.\/} {\bf 117}(12) 123604 (\textit{Preprint} \eprint{1605.03990})

\bibitem{Tongcang_rotation:2018}
Ahn J, Xu Z, Bang J, Deng Y~H, Hoang T~M, Han Q, Ma R~M and Li T 2018 {\em
  Phys. Rev. Lett.\/} {\bf 121}(3) 033603

\bibitem{Millen2017_2}
Kuhn S, Stickler B~A, Kosloff A, Patolsky F, Hornberger K, Arndt M and Millen J
  2017 {\em Nat. Commun.\/} {\bf 8} 1670

\bibitem{vanderLaan2020}
van~der Laan F, Reimann R, Militaru A, Tebbenjohanns F, Windey D, Frimmer M and
  Novotny L 2020 {\em Phys. Rev. A\/} {\bf 102}(1) 013505

\bibitem{Blakemore_3D_microscope:2019}
Blakemore C~P, Rider A~D, Roy S, Wang Q, Kawasaki A and Gratta G 2019 {\em
  Phys. Rev. A\/} {\bf 99}(2) 023816

\bibitem{Hempston:2017}
Hempston D, Vovrosh J, Toro M, Winstone G, Rashid M and Ulbricht H 2017 {\em
  Appl. Phys. Lett.\/} {\bf 111} 133111 (\textit{Preprint} \eprint{1706.09774})

\bibitem{blakemore_gauge:2019}
Blakemore C~P, Martin D, Fieguth A, Kawasaki A, Priel N, Rider A~D and Gratta G
  2019  (\textit{Preprint} \eprint{1911.09090})

\bibitem{EotWash2020}
Lee J~G, Adelberger E~G, Cook T~S, Fleischer S~M and Heckel B~R 2020 {\em Phys.
  Rev. Lett.\/} {\bf 124}(10) 101101

\bibitem{Asenbaum:2020era}
Asenbaum P, Overstreet C, Kim M, Curti J and Kasevich M~A 2020
  (\textit{Preprint} \eprint{2005.11624})

\bibitem{Gilmore:2017cbw}
Gilmore K, Bohnet J, Sawyer B, Britton J and Bollinger J 2017 {\em Phys. Rev.
  Lett.\/} {\bf 118} 263602 (\textit{Preprint} \eprint{1703.05369})

\bibitem{Biercuk:2010}
{Biercuk} M~J, {Uys} H, {Britton} J~W, {VanDevender} A~P and {Bollinger} J~J
  2010 {\em arXiv e-prints\/} arXiv:1004.0780 (\textit{Preprint}
  \eprint{1004.0780})

\bibitem{Carney_white_paper}
Carney D {\em et~al.\/} 2020 {Mechanical Quantum Sensing in the Search for Dark
  Matter} (\textit{Preprint} \eprint{2008.06074})

\bibitem{carney:2019_1}
Carney D, Ghosh S, Krnjaic G and Taylor J~M 2019  (\textit{Preprint}
  \eprint{1903.00492})

\bibitem{2000AAMOP..42...95G}
{Grimm} R, {Weidem{\"u}ller} M and {Ovchinnikov} Y~B 2000 {\em Advances in
  Atomic Molecular and Optical Physics\/} {\bf 42} 95--170 (\textit{Preprint}
  \eprint{physics/9902072})

\bibitem{Frimmer:2017}
{Frimmer} M, {Luszcz} K, {Ferreiro} S, {Jain} V, {Hebestreit} E and {Novotny} L
  2017 {\em Phys. Rev. A\/} {\bf 95} 061801 (\textit{Preprint}
  \eprint{1704.00169})

\bibitem{Conangla:2018nnn}
Conangla G~P, Ricci F, Cuairan M~T, Schell A~W, Meyer N and Quidant R 2019 {\em
  Phys. Rev. Lett.\/} {\bf 122} 223602 (\textit{Preprint} \eprint{1901.00923})

\bibitem{Chang:2010}
{Chang} D~E, {Regal} C~A, {Papp} S~B, {Wilson} D~J, {Ye} J, {Painter} O,
  {Kimble} H~J and {Zoller} P 2010 {\em Proc. Nat. Acad. Sci. USA\/} {\bf 107}
  1005--1010 (\textit{Preprint} \eprint{0909.1548})

\bibitem{Jain:2016}
Jain V, Gieseler J, Moritz C, Dellago C, Quidant R and Novotny L 2016 {\em
  Phys. Rev. Lett.\/} {\bf 116}(24) 243601 (\textit{Preprint}
  \eprint{1603.03420})

\bibitem{Aggarwal:2020umq}
Aggarwal N, Winstone G~P, Teo M, Baryakhtar M, Larson S~L, Kalogera V and
  Geraci A~A 2020  (\textit{Preprint} \eprint{2010.13157})

\bibitem{Chang_2012}
Chang D~E, Ni K~K, Painter O and Kimble H~J 2012 {\em New Journal of Physics\/}
  {\bf 14} 045002

\bibitem{2013Sci...339..801P}
{Purdy} T~P, {Peterson} R~W and {Regal} C~A 2013 {\em Science\/} {\bf 339}
  801--804 (\textit{Preprint} \eprint{1209.6334})

\bibitem{PhysRevD.23.1693}
Caves C~M 1981 {\em Phys. Rev. D\/} {\bf 23}(8) 1693--1708
  \urlprefix\url{https://link.aps.org/doi/10.1103/PhysRevD.23.1693}

\bibitem{Tebben2019}
Tebbenjohanns F, Frimmer M and Novotny L 2019 {\em Phys. Rev. A\/} {\bf 100}(4)
  043821

\bibitem{Mason:2019piu}
Mason D, Chen J, Rossi M, Tsaturyan Y and Schliesser A 2019 {\em Nature
  Phys.\/} {\bf 15} 745--749 (\textit{Preprint} \eprint{1809.10629})

\bibitem{carney:2019_2}
Carney D, Hook A, Liu Z, Taylor J~M and Zhao Y 2019  (\textit{Preprint}
  \eprint{1908.04797})

\bibitem{Kubo_1966}
Kubo R 1966 {\em Rep. Prog. Phys.\/} {\bf 29} 255--284

\bibitem{Novotny:2020}
Tebbenjohanns F, Frimmer M, Jain V, Windey D and Novotny L 2020 {\em Phys. Rev.
  Lett.\/} {\bf 124}(1) 013603

\bibitem{Novotny2019}
Tebbenjohanns F, Frimmer M, Militaru A, Jain V and Novotny L 2019 {\em Phys.
  Rev. Lett.\/} {\bf 122}(22) 223601

\bibitem{Rider:2017}
Rider A~D, Blakemore C~P, Gratta G and Moore D~C 2018 {\em Phys. Rev. A\/} {\bf
  97}(1) 013842

\bibitem{Kawasaki:2020oyl}
Kawasaki A, Fieguth A, Priel N, Blakemore C~P, Martin D and Gratta G 2020
  (\textit{Preprint} \eprint{2004.10973})

\bibitem{Aspelmeyer:2018}
Belenchia A, Wald R~M, Giacomini F, Castro-Ruiz E, Brukner C and Aspelmeyer M
  2018 {\em Phys. Rev. D\/} {\bf 98}(12) 126009

\bibitem{arita2016rotational}
Arita Y, Richards J~M, Mazilu M, Spalding G~C, Skelton~Spesyvtseva S~E, Craig D
  and Dholakia K 2016 {\em ACS Nano.\/} {\bf 10} 11505--11510

\bibitem{Rotation_paper_2018}
Monteiro F, Ghosh S, van Assendelft E~C and Moore D~C 2018 {\em Phys. Rev. A\/}
  {\bf 97}(5) 051802

\bibitem{Rider:2019}
Rider A~D, Blakemore C~P, Kawasaki A, Priel N, Roy S and Gratta G 2019 {\em
  Phys. Rev. A\/} {\bf 99}(4) 041802

\bibitem{Ahn:2020}
{Ahn} J, {Xu} Z, {Bang} J, {Ju} P, {Gao} X and {Li} T 2020 {\em Nature
  Nanotech.\/} {\bf 15} 89--93 (\textit{Preprint} \eprint{1908.03453})

\bibitem{Oriol2011}
Romero-Isart O, Pflanzer A~C, Blaser F, Kaltenbaek R, Kiesel N, Aspelmeyer M
  and Cirac J~I 2011 {\em Phys. Rev. Lett.\/} {\bf 107}(2) 020405

\bibitem{Geraci:matter_wave}
Geraci A and Goldman H 2015 {\em Phys. Rev. D\/} {\bf 92}(6) 062002

\bibitem{Ulbricht:2014}
Bateman J, Nimmrichter S, Hornberger K and Ulbricht H 2014 {\em Nature
  Communications\/} {\bf 5} 4788

\bibitem{Marletto:2017}
Marletto C and Vedral V 2017 {\em Phys. Rev. Lett.\/} {\bf 119}(24) 240402

\bibitem{Bose:2017}
Bose S, Mazumdar A, Morley G~W, Ulbricht H, Toro\ifmmode~\check{s}\else
  \v{s}\fi{} M, Paternostro M, Geraci A~A, Barker P~F, Kim M~S and Milburn G
  2017 {\em Phys. Rev. Lett.\/} {\bf 119}(24) 240401

\bibitem{Monteiro:2020wcb}
Monteiro F, Afek G, Carney D, Krnjaic G, Wang J and Moore D~C 2020
  (\textit{Preprint} \eprint{2007.12067})

\bibitem{matterwavereview}
Cronin A~D, Schmiedmayer J and Pritchard D~E 2009 {\em Rev. Mod. Phys.\/} {\bf
  81}(3) 1051--1129

\bibitem{arndt2019}
Fein Y~Y, Geyer P, Zwick P, Kia{\l}ka F, Pedalino S, Mayor M, Gerlich S and
  Arndt M 2019 {\em Nature Physics\/} {\bf 15} 1242--1245

\bibitem{Essig:2013lka}
Essig R {\em et~al.\/} 2013 {Working Group Report: New Light Weakly Coupled
  Particles} {\em {Community Summer Study 2013}: {Snowmass on the
  Mississippi}\/} (\textit{Preprint} \eprint{1311.0029})

\bibitem{Murata_2015}
Murata J and Tanaka S 2015 {\em Class. Quant. Grav.\/} {\bf 32} 033001

\bibitem{Lamoreaux:2012}
Lamoreaux S~K 2012 {\em Ann. Rev. Nucl. Part. Sci.\/} {\bf 62} 37--56

\bibitem{Geraci:2010}
Geraci A~A, Papp S~B and Kitching J 2010 {\em Phys. Rev. Lett.\/} {\bf 105}(10)
  101101 (\textit{Preprint} \eprint{1006.0261})

\bibitem{Onofrio:2006mq}
Onofrio R 2006 {\em New J. Phys.\/} {\bf 8} 237 (\textit{Preprint}
  \eprint{hep-ph/0612234})

\bibitem{Xu:2017bpr}
Xu Z and Li T 2017 {\em Phys. Rev. A\/} {\bf 96} 033843 (\textit{Preprint}
  \eprint{1704.08770})

\bibitem{Garrett:2020}
Garrett J~L, Kim J and Munday J~N 2020 {\em Phys. Rev. Research\/} {\bf 2}(2)
  023355

\bibitem{Garrett:2015bde}
Garrett J~L, Somers D and Munday J~N 2015 {\em J. Phys. Condens. Matter\/} {\bf
  27} 214012 (\textit{Preprint} \eprint{1409.5012})

\bibitem{kim2008anomalies}
Kim W, Brown-Hayes M, Dalvit D, Brownell J and Onofrio R 2008 {\em Phys. Rev.
  A\/} {\bf 78} 020101

\bibitem{Rene:2018}
Reimann R, Doderer M, Hebestreit E, Diehl R, Frimmer M, Windey D, Tebbenjohanns
  F and Novotny L 2018 {\em Phys. Rev. Lett.\/} {\bf 121}(3) 033602

\bibitem{Moore:2018spie}
{Moore} D~C 2018 {\em Proc. SPIE, Optical Trapping and Optical
  Micromanipulation XV\/}  107230H

\bibitem{Rider:2016}
Rider A~D, Moore D~C, Blakemore C~P, Louis M, Lu M and Gratta G 2016 {\em Phys.
  Rev. Lett.\/} {\bf 117}(10) 101101 (\textit{Preprint} \eprint{1604.04908})

\bibitem{Wang:2017}
{Wang} Q, {Rider} A~D, {Moore} D~C, {Blakemore} C~P, {Cao} L and {Gratta} G
  2017 A density staggered cantilever for micron length gravity probing {\em
  2017 IEEE 67th Electronic Components and Technology Conference (ECTC)\/} pp
  1773--1778

\bibitem{Schmole:2016mde}
Schm\"ole J, Dragosits M, Hepach H and Aspelmeyer M 2016 {\em Class. Quant.
  Grav.\/} {\bf 33} 125031 (\textit{Preprint} \eprint{1602.07539})

\bibitem{Westphal:2020okx}
Westphal T, Hepach H, Pfaff J and Aspelmeyer M 2020  (\textit{Preprint}
  \eprint{2009.09546})

\bibitem{PhysRevD.68.124021}
Dimopoulos S and Geraci A~A 2003 {\em Phys. Rev. D\/} {\bf 68}(12) 124021
  \urlprefix\url{https://link.aps.org/doi/10.1103/PhysRevD.68.124021}

\bibitem{Kovachy:2015}
Kovachy T, Asenbaum P, Overstreet C, Donnelly C~A, Dickerson S~M, Sugarbaker A,
  Hogan J~M and Kasevich M~A 2015 {\em Nature\/} {\bf 528} 530--533

\bibitem{Jaeckel:2009dh}
Jaeckel J 2009 {\em Phys. Rev. Lett.\/} {\bf 103} 080402 (\textit{Preprint}
  \eprint{0904.1547})

\bibitem{Jaeckel:2010}
Jaeckel J and Roy S 2010 {\em Phys. Rev. D\/} {\bf 82}(12) 125020

\bibitem{HOLDOM:1986}
Holdom B 1986 {\em Physics Letters B\/} {\bf 166} 196 -- 198

\bibitem{EHRET2010149}
Ehret K, Frede M, Ghazaryan S, Hildebrandt M, Knabbe E~A, Kracht D, Lindner A,
  List J, Meier T, Meyer N, Notz D, Redondo J, Ringwald A, Wiedemann G and
  Willke B 2010 {\em Physics Letters B\/} {\bf 689} 149 -- 155

\bibitem{Unnikrishnan_2004}
Unnikrishnan C~S and Gillies G~T 2004 {\em Metrologia\/} {\bf 41} S125--S135

\bibitem{Foot:1993yq}
Foot R 1994 {\em Phys. Rev. D\/} {\bf 49} 3617--3621 (\textit{Preprint}
  \eprint{hep-ph/9402240})

\bibitem{Ball:2020dnx}
Ball A {\em et~al.\/} 2020 {\em Phys. Rev. D\/} {\bf 102} 032002
  (\textit{Preprint} \eprint{2005.06518})

\bibitem{Prinz:1998ua}
Prinz A {\em et~al.\/} 1998 {\em Phys. Rev. Lett.\/} {\bf 81} 1175--1178
  (\textit{Preprint} \eprint{hep-ex/9804008})

\bibitem{Baumann:1988}
Baumann J, G\"ahler R, Kalus J and Mampe W 1988 {\em Phys. Rev. D\/} {\bf
  37}(11) 3107--3112

\bibitem{Bressi:2011}
Bressi G, Carugno G, Della~Valle F, Galeazzi G, Ruoso G and Sartori G 2011 {\em
  Phys. Rev. A\/} {\bf 83}(5) 052101

\bibitem{Morpurgo:1984}
Marinelli M and Morpurgo G 1984 {\em Phys. Lett. B\/} {\bf 137} 439 -- 442

\bibitem{PhysRevLett.99.161804}
Kim P~C, Lee E~R, Lee I~T, Perl M~L, Halyo V and Loomba D 2007 {\em Phys. Rev.
  Lett.\/} {\bf 99}(16) 161804

\bibitem{Ashkin1980}
Ashkin A 1980 {\em Science\/} {\bf 210} 1081--1088

\bibitem{atom2008}
Arvanitaki A, Dimopoulos S, Geraci A~A, Hogan J and Kasevich M 2008 {\em Phys.
  Rev. Lett.\/} {\bf 100}(12) 120407

\bibitem{axion1}
Peccei R~D and Quinn H~R 1977 {\em Phys. Rev. Lett.\/} {\bf 38}(25) 1440--1443

\bibitem{axion2}
Weinberg S 1978 {\em Phys. Rev. Lett.\/} {\bf 40}(4) 223--226

\bibitem{PTViolation}
Wilczek F 1978 {\em Phys. Rev. Lett.\/} {\bf 40}(5) 279--282

\bibitem{Moody:1984ba}
Moody J~E and Wilczek F 1984 {\em Phys. Rev. D\/} {\bf 30}(1) 130--138

\bibitem{stringaxiverse}
Arvanitaki A, Dimopoulos S, Dubovsky S, Kaloper N and March-Russell J 2010 {\em
  Phys. Rev. D\/} {\bf 81}(12) 123530

\bibitem{stringaxiverse2}
Arvanitaki A and Dubovsky S 2011 {\em Phys. Rev. D\/} {\bf 83}(4) 044026

\bibitem{minamasha2014}
Arvanitaki A, Baryakhtar M and Huang X 2015 {\em Phys. Rev. D\/} {\bf 91}(8)
  084011 \urlprefix\url{https://link.aps.org/doi/10.1103/PhysRevD.91.084011}

\bibitem{GWprl}
Arvanitaki A and Geraci A~A 2013 {\em Phys. Rev. Lett.\/} {\bf 110}(7) 071105

\bibitem{GWlev2020}
Aggarwal N, Winstone G, Baryakhtar M, Teo M, S L, Kalogera V and Geraci A 2020
  {\em in preparation\/}

\bibitem{GWMoore_2014}
Moore C~J, Cole R~H and Berry C~P~L 2014 {\em Classical and Quantum Gravity\/}
  {\bf 32} 015014

\bibitem{SougatoGW}
Marshman R~J, Mazumdar A, Morley G~W, Barker P~F, Hoekstra S and Bose S 2020
  {\em New Journal of Physics\/} {\bf 22} 083012

\bibitem{Schumann:2019eaa}
Schumann M 2019 {\em J. Phys. G\/} {\bf 46} 103003 (\textit{Preprint}
  \eprint{1903.03026})

\bibitem{Graham:2015ouw}
Graham P~W, Irastorza I~G, Lamoreaux S~K, Lindner A and van Bibber K~A 2015
  {\em Ann. Rev. Nucl. Part. Sci.\/} {\bf 65} 485--514 (\textit{Preprint}
  \eprint{1602.00039})

\bibitem{Knapen:2017ekk}
Knapen S, Lin T, Pyle M and Zurek K~M 2018 {\em Phys. Lett. B\/} {\bf 785}
  386--390 (\textit{Preprint} \eprint{1712.06598})

\bibitem{Coskuner:2019}
Coskuner A, Grabowska D~M, Knapen S and Zurek K~M 2019 {\em Phys. Rev. D\/}
  {\bf 100}(3) 035025

\bibitem{Griffin:2018bjn}
Griffin S, Knapen S, Lin T and Zurek K~M 2018 {\em Phys. Rev. D\/} {\bf 98}
  115034 (\textit{Preprint} \eprint{1807.10291})

\bibitem{Carney:2020xol}
Carney D {\em et~al.\/} 2020 {Mechanical Quantum Sensing in the Search for Dark
  Matter} (\textit{Preprint} \eprint{2008.06074})

\bibitem{Hall:2018}
Hall E~D, Adhikari R~X, Frolov V~V, M\"uller H and Pospelov M 2018 {\em Phys.
  Rev. D\/} {\bf 98}(8) 083019

\bibitem{AkioDMpaper}
Kawasaki A 2019 {\em Phys. Rev. D\/} {\bf 99}(2) 023005

\bibitem{Cheng:2019vwy}
Cheng T, Primulando R and Spinrath M 2020 {\em Eur. Phys. J. C\/} {\bf 80} 519
  (\textit{Preprint} \eprint{1906.07356})

\bibitem{Lee:2020dcd}
Lee C~H, Nugroho C~S and Spinrath M 2020  (\textit{Preprint}
  \eprint{2007.07908})

\bibitem{Bateman:2014lia}
Bateman J, McHardy I, Merle A, Morris T~R and Ulbricht H 2015 {\em Sci. Rep.\/}
  {\bf 5} 8058 (\textit{Preprint} \eprint{1405.5536})

\bibitem{Riedel:2016acj}
Riedel C~J and Yavin I 2017 {\em Phys. Rev. D\/} {\bf 96} 023007
  (\textit{Preprint} \eprint{1609.04145})

\bibitem{Ahlen:2009ev}
Ahlen S {\em et~al.\/} 2010 {\em Int. J. Mod. Phys. A\/} {\bf 25} 1--51
  (\textit{Preprint} \eprint{0911.0323})

\bibitem{2020arXiv200501948W}
{Wang} J, {Sciarrino} F, {Laing} A and {Thompson} M~G 2020 {\em arXiv
  e-prints\/} arXiv:2005.01948 (\textit{Preprint} \eprint{2005.01948})

\bibitem{2016Sci...354.1021B}
{Barredo} D, {de L{\'e}s{\'e}leuc} S, {Lienhard} V, {Lahaye} T and {Browaeys} A
  2016 {\em Science\/} {\bf 354} 1021--1023 (\textit{Preprint}
  \eprint{1607.03042})

\bibitem{2016Sci...354.1024E}
{Endres} M, {Bernien} H, {Keesling} A, {Levine} H, {Anschuetz} E~R,
  {Krajenbrink} A, {Senko} C, {Vuletic} V, {Greiner} M and {Lukin} M~D 2016
  {\em Science\/} {\bf 354} 1024--1027

\bibitem{BassiRMP}
Bassi A, Lochan K, Satin S, Singh T~P and Ulbricht H 2013 {\em Rev. Mod.
  Phys.\/} {\bf 85}(2) 471--527

\bibitem{Penrose1996}
Penrose R 1996 {\em General Relativity and Gravitation\/} {\bf 28} 581--600

\bibitem{Bassi:2017}
Bassi A, Gro{\ss}ardt A and Ulbricht H 2017 {\em Classical and Quantum
  Gravity\/} {\bf 34} 193002

\bibitem{Diosi1989}
Di\'osi L 1989 {\em Phys. Rev. A\/} {\bf 40}(3) 1165--1174

\bibitem{Carlesso:19}
Carlesso M and Bassi A 2019 Current tests of collapse models: How far can we
  push the limits of quantum mechanics? {\em Quantum Information and
  Measurement (QIM) V: Quantum Technologies\/} (Optical Society of America) p
  S1C.3

\bibitem{Zheng2020}
Zheng D, Leng Y, Kong X, Li R, Wang Z, Luo X, Zhao J, Duan C~K, Huang P, Du J,
  Carlesso M and Bassi A 2020 {\em Phys. Rev. Research\/} {\bf 2}(1) 013057
  \urlprefix\url{https://link.aps.org/doi/10.1103/PhysRevResearch.2.013057}

\bibitem{Vinante2020}
Vinante A, Gasbarri G, Timberlake C, Toro\ifmmode~\check{s}\else \v{s}\fi{} M
  and Ulbricht H 2020 {\em Phys. Rev. Research\/} {\bf 2}(4) 043229
  \urlprefix\url{https://link.aps.org/doi/10.1103/PhysRevResearch.2.043229}

\bibitem{VinantePRL2020}
Vinante A, Carlesso M, Bassi A, Chiasera A, Varas S, Falferi P, Margesin B,
  Mezzena R and Ulbricht H 2020 {\em Phys. Rev. Lett.\/} {\bf 125}(10) 100404
  \urlprefix\url{https://link.aps.org/doi/10.1103/PhysRevLett.125.100404}

\bibitem{Donadi2020}
Donadi S, Piscicchia K, Curceanu C, Di{\'o}si L, Laubenstein M and Bassi A 2020
  {\em Nature Physics\/} ISSN 1745-2481
  \urlprefix\url{https://doi.org/10.1038/s41567-020-1008-4}

\bibitem{Rahman2017}
Rahman A~T~M~A and Barker P~F 2017 {\em Nature Photonics\/} {\bf 11} 634--638

\bibitem{Oriolracetrack}
Pino H, Prat-Camps J, Sinha K, Venkatesh B~P and Romero-Isart O 2018 {\em
  Quantum Science and Technology\/} {\bf 3} 025001

\bibitem{PhysRevLett.47.979}
Page D~N and Geilker C~D 1981 {\em Phys. Rev. Lett.\/} {\bf 47}(14) 979--982
  \urlprefix\url{https://link.aps.org/doi/10.1103/PhysRevLett.47.979}

\bibitem{2013arXiv1311.4558K}
{Kafri} D and {Taylor} J~M 2013 {\em arXiv e-prints\/} arXiv:1311.4558
  (\textit{Preprint} \eprint{1311.4558})

\bibitem{Carlesso:2019}
Carlesso M, Bassi A, Paternostro M and Ulbricht H 2019 {\em New Phys.\/} {\bf
  21} 093052

\bibitem{Carlesso:2017vrw}
Carlesso M, Paternostro M, Ulbricht H and Bassi A 2017  (\textit{Preprint}
  \eprint{1710.08695})

\end{thebibliography}

\end{document}